\documentclass[12pt]{article}

\usepackage{cite}
\usepackage{epsfig}
\usepackage{amsmath,amsfonts}
\usepackage{url}

\newcommand{\mst}[1]{m_{\tilde{t_{#1}}}}

\newcommand{\st}{{\tilde{t}}}
\newcommand{\sB}{{\tilde{b}}}

\newcommand{\glo}{{\tilde{g}}}

\newcommand{\mseR}{m_{\tilde{e}_{\rm R}}}

\newcommand{\cha}{\tilde{\chi}}
\newcommand{\neu}{\tilde{\chi}^0}
\newcommand{\mcha}[1]{m_{\tilde{\chi}^\pm_{#1}}}
\newcommand{\mneu}[1]{m_{\tilde{\chi}^0_{#1}}}

\newcommand{\NPB}[3]{{ Nucl.~Phys.} \textbf{B#1}, #3 (#2)}   
\newcommand{\PLB}[3]{{ Phys.~Lett.} \textbf{B#1}, #3 (#2)}   
\newcommand{\PRD}[3]{{ Phys.~Rev.} \textbf{D#1}, #3 (#2)}

\newcommand{\HPA}[3]{{ Helv.~Phys.~Acta} \textbf{#1}, #3 (#2)}

\newcommand{\ms}{m_{\rm s}}
\newcommand{\tad}{t_{\rm s}}
\newcommand{\al}{a_\lambda}
\newcommand{\vs}{v_{\rm s}}

\def\mathswitch#1{\relax\ifmmode#1\else$#1$\fi}
\def\mathswitchr#1{\relax\ifmmode{\mathrm{#1}}\else$\mathrm{#1}$\fi}
\newcommand{\PW}{\mathswitchr W}
\newcommand{\PZ}{\mathswitchr Z}

\newcommand{\Pt}{\mathswitchr t}
\newcommand{\PA}{\mathswitchr A}
\newcommand{\MW}{\mathswitch {M_\PW}}
\newcommand{\MZ}{\mathswitch {M_\PZ}}

\newcommand{\MA}{\mathswitch {M_\PA}}

\newcommand{\mt}{\mathswitch {m_\Pt}}

\newcommand{\scrs}{{}}
\newcommand{\sw}{\mathswitch {s_{\scrs\PW}}}
\newcommand{\cw}{\mathswitch {c_{\scrs\PW}}}

\newcommand{\tev}{\,\, \mathrm{TeV}}
\newcommand{\gev}{\,\, \mathrm{GeV}}
\newcommand{\mev}{\,\, \mathrm{MeV}}
\newcommand{\re}{\Re e \,}
\newcommand{\im}{\Im m \,}

\newcommand{\SLASH}[2]{\makebox[#2ex][l]{$#1$}/}

\newcommand{\pslash}{\SLASH{p}{.2}}
\newcommand{\ppslash}{\SLASH{\vec{p}}{.2}}

\newcommand{\Eslash}{\SLASH{E}{.5}\,}
\newcommand{\RR}{\text{R}}
\newcommand{\LL}{\text{L}}

\newcommand{\anc}{\rule{0mm}{0mm}}
\newcommand{\lesim}{\,\raisebox{-.1ex}{$_{\textstyle <}\atop^{\textstyle\sim}$}\,}
\newcommand{\gesim}{\,\raisebox{-.3ex}{$_{\textstyle >}\atop^{\textstyle\sim}$}\,}

\newcommand{\mycaption}[1]{\caption{\sl #1}}

\begin{document}

\title{
\vspace{-2em}
\begin{flushright}
\normalsize ANL-HEP-PR-07-22 \\
\normalsize EFI-07-09 \\
\normalsize FERMILAB-PUB-07-080-T \\
\normalsize ZU-TH 11/07 \\
\end{flushright}
\vskip 30pt
Phenomenology of the nMSSM \\ from colliders to cosmology
}

\author{
C. Bal\'azs$^{1,2,3}$, M. Carena$^{3}$, A. Freitas$^{4}$ and C.E.M. Wagner$^{2,5}$
\\[1em] 
\small \sl $^{1}$ School of Physics, Monash University, Melbourne VIC 3800,
Australia \\
\small \sl $^{2}$ HEP Division, Argonne National Laboratory, 9700 Cass Ave.,
Argonne, IL 60439, USA \\
\small \sl $^{3}$ Fermi National Accelerator Laboratory, P.O.~Box 500, Batavia,
IL 60510, USA \\
\small \sl $^{4}$ Institut f\"ur Theoretische Physik, Universit\"at Z\"urich,
 Winterthurerstrasse 190, \\[-1ex] \small \sl 8057 Z\"urich, Switzerland\\
\small \sl $^{5}$ Enrico Fermi Institute and Kavli Institute for
Cosmological Physics,\\[-1ex]  \small \sl  Department of Physics, University of
Chicago, 5640 S. Ellis
Ave.,\\[-1ex]  \small \sl  
Chicago, IL 60637, USA }

\date{}

\maketitle

\begin{abstract}

Low energy supersymmetric models provide a solution to the hierarchy problem 
and also have the necessary ingredients to solve two of the most outstanding
issues in cosmology: the origin of dark matter and baryonic matter. One of the
most attractive features of this framework is that the relevant physical
processes are related to interactions at the weak scale and therefore may be
tested  in collider experiments in the near future. This is true for the
Minimal  Supersymmetric Standard Model (MSSM) as well as for its extension
with  the addition of one singlet chiral superfield, the so-called nMSSM. It
has been recently shown that within the nMSSM an elegant solution to both the
problem of baryogenesis and  dark matter may be found, that relies mostly on
the mixing of the singlet sector with the Higgs sector of the theory. In this
work we review the nMSSM model constraints from cosmology and present the
associated collider phenomenology at the LHC and the ILC. We show that the ILC
will efficiently probe the neutralino, chargino and Higgs sectors, allowing to
confront cosmological observations with computations based on collider
measurements. We also investigate the  prospects for a direct detection of dark
matter and the constraints imposed by the current bounds of the electron
electric dipole moment in this model. 

\end{abstract}

\thispagestyle{empty}

\setcounter{page}{0}
\setcounter{footnote}{0}

\newpage

%%%%%%%%%%%%%%%%%%%%%%%%%%%%%%%%%%%%%%%%%%%%%%%%%%%%%%%%%%%%%%
%%%%%%%%%%%%%%%%%%%%%%%%%%%%%%%%%%%%%%%%%%%%%%%%%%%%%%%%%%%%%%

\section{Introduction}

In spite of the excellent agreement of the predictions of the 
Standard Model (SM) with experimental observables, there is
still a strong physical motivation for the presence of new physics
at the weak scale. The main reason is the belief that the
Higgs mechanism of the SM is only an effective description of
a more fundamental mechanism of electroweak symmetry breaking,
in which the origin and stability of the electroweak scale must
be explained.  Supersymmetric extensions of the 
SM~\cite{Haber:1984rc,Martin:1997ns} 
allow such a mechanism.
In the simplest extensions of the SM, the theory remains perturbative
up to scales of the order of the Planck scale and the weak scale
is stable under quantum corrections at all orders in perturbation 
theory. Moreover, the electroweak symmetry is
broken radiatively,  providing a correlation between the weak scale 
and the soft supersymmetry breaking scale.

Another strong motivation is a solution to the problem 
of dark matter. If a discrete symmetry, R-Parity, is
imposed in the theory, the lightest supersymmetric particle (LSP) is
stable. In the simplest models, it is also neutral and weakly
interacting, with an annihilation cross section of the order of
the one necessary for the LSP to become a good dark matter candidate.

It has been also realized that supersymmetry may also lead
to a solution of another outstanding cosmological problem, namely,
the baryon-antibaryon asymmetry~\cite{EWBGreviews,PT}. 
In the MSSM, a solution to 
this problem by weak scale physics
demands a light stop as well as non-vanishing CP-violating
phases in the chargino--neutralino sector. The presence of a light
stop, with mass smaller than the top quark mass, tends to push the
lightest CP-even Higgs mass to values close to the ones restricted
by experimental searches. Therefore, this scenario is highly constrained
and it will be probed at the Tevatron and the LHC colliders in the
near future, as well as in electric dipole moment and direct dark
matter searches~\cite{Csaba,Caltech} and the ILC~\cite{stop}.  
Supersymmetry also plays
a relevant role for the generation of the baryon asymmetry in the 
so-called soft leptogenesis scenario, which again demands very
specific values for the soft supersymmetry breaking parameters in
the neutrino sector~\cite{softlepto}, which can be naturally obtained
only in certain supersymmetry breaking scenarios~\cite{naturalsoftl}.

A less constrained solution to the baryon asymmetry
may be found within extensions of the MSSM
with additional chiral singlet superfields. In particular, it has
been recently shown that a particularly simple extension of the
MSSM, the so-called nMSSM~\cite{xmssm,nMSSM,tadloop,hcorr,nMSSM2} 
includes all the relevant properties
to lead to a successful generation of the baryon asymmetry at the
weak scale~\cite{Kang:2004pp,gaga}, 
without the need of light squarks and with values of
the lightest CP-even Higgs mass which are naturally above the 
current experimental bound~\cite{lephbound}.

One interesting feature of the nMSSM is that the lightest neutralino is usually
an admixture of the fermion component of the singlet field and the Higgsinos.
Due to the absence of explicit mass terms in the Higgsino--singlino sector, its
mass is bounded to be below 70 GeV. A relaxation of this upper bound may only
be obtained by allowing  the trilinear Higgs-singlet coupling to become strong
at scales lower than the Grand Unification (GUT) scale, as happens in the
so-called fat Higgs models and gauge extensions\footnote{Such gauge extensions
open new possibilities for baryogenesis~\cite{gaugebar}, but this possibility
will not be pursued any further in this paper.} of the
nMSSM~\cite{Harnik:2003rs}.  In this work, we shall demand perturbative
consistency of the theory up to the GUT scale, and therefore the aforementioned
bound applies. For light neutralinos, the annihilation cross section is
dominated by s-channel $Z^0$ boson exchange and a neutralino relic density
consistent with all experimental bounds, may  only be obtained for values of
the neutralino mass of about 30--45 GeV.

Such a light neutralino has relevant phenomenological consequences. 
For instance, the lightest CP-even Higgs tends to decay predominantly
into neutralinos, leading to invisible signatures at hadron and
lepton colliders~\cite{gaga}. Moreover, one of the predictions of this model
is the presence of relatively light charginos and neutralinos with
masses smaller than about 200 GeV. Detection of these weakly interacting
particles at hadron colliders becomes difficult if there are no 
strongly interacting superpartners with masses below the TeV scale. 
Assuming that the relevant CP-violating phases are in the gaugino
masses, successful baryogenesis may be easily achieved for weak-scale
gaugino masses of the order of 100--300 GeV, and at least one
relatively light squark in order to avoid the so-called Giudice-Shaposhnikov
suppression of chiral charges~\cite{GS}. 
We will assume the presence of
top and bottom squarks with masses of order of 500 GeV, and 
the presence of a gluino with mass dictated by the gaugino mass
unification condition. The phenomenological properties at hadron
colliders depend strongly on these last assumptions.

In addition to discussing the prospects for discovery of the light Higgs and
superpartner states at the LHC, we will perform a detailed
phenomenological analysis of nMSSM physics at the ILC. 
In this case the values of the strongly interacting
particle masses become less important. The same is true for
the analysis of the direct dark matter detection properties
and the constraints coming from the bounds on the electron 
electric dipole moments that we investigate in this work.

The article is organized as follows. In section \ref{sc:nmssm} we present an
overview of the nMSSM and present experimental constraints to define a benchmark 
scenario for the model. In section \ref{sc:coll} we analyze the corresponding
collider phenomenology at the LHC and the ILC, which we then connect with
the cosmology constraints related to the dark
matter relic density and baryogenesis
in section \ref{sc:cosmo}. In particular, we present
the constraints coming from direct searches of dark matter and
the non-observation of an electron electric dipole moment. We
reserve section \ref{sc:concl} for our conclusions.

%%%%%%%%%%%%%%%%%%%%%%%%%%%%%%%%%%%%%%%%%%%%%%%%%%%%%%%%%%%%%%

\section{General properties of and constraints on the nMSSM}
\label{sc:nmssm}

\subsection{Overview}

One of the original motivations for a singlet extension of the MSSM is the
difficulty to explain why the $\mu$-parameter in the superpotential is of order
of the electroweak scale, instead of the GUT or Planck scale.
This problem is circumvented by introducing a singlet
chiral superfield $\hat{S}$, and generating a
weak-scale $\mu$-parameter through the vacuum expectation value of the scalar
component $S$ of the singlet. This requires the inclusion of a triple-Higgs
coupling term 
\begin{equation} 
W_{\lambda} = \lambda \hat{S} \hat{H}_1 \cdot \hat{H}_2 
\end{equation}
in the
superpotential, where $\hat{H}_{1,2}$ are the superfields of the two Higgs
doublets,
\begin{align}
\hat{H}_1 &= \begin{pmatrix} H_1^0 \\ H_1^- \end{pmatrix}, &
\hat{H}_2 &= \begin{pmatrix} H_2^+ \\ H_2^0 \end{pmatrix}.
\end{align}
This triple-Higgs coupling also helps to push up the mass of
the CP-even Higgs boson responsible for electroweak symmetry breaking, so that
the bound from LEP, \linebreak
 $m_{\rm h^0} \gesim 114.4 \gev$~\cite{lephbound}, can be
satisfied without substantial fine tuning. Finally, the new contribution
increases the strength of the electroweak phase transition, thus allowing the
possibility of electroweak baryogenesis in a large region of parameter space
\cite{gaga}.

In a general singlet extension, the superpotential can also contain a
triple-self-coupling, a mass term and a tadpole term for $\hat{S}$,
\begin{equation}
\begin{aligned}
W_{\rm SMSSM} &= \lambda \hat{S} \hat{H}_1 \cdot \hat{H}_2 + \kappa \hat{S}^3 + 
  m_{\rm N} \hat{S}^2 +
  \frac{m^2_{12}}{\lambda}  \hat{S} \\
   & \;\;\;\; + y_{\rm u} \hat{Q} \cdot \hat{H}_2 \, \hat{U}^c 
         + y_{\rm d} \hat{Q} \cdot \hat{H}_1 \, \hat{D}^c
         + y_{\rm e} \hat{L} \cdot \hat{H}_1 \, \hat{E}^c,
\end{aligned}
\end{equation}
where $y_{\rm u,d,e}$ are the $3\times 3$ Yukawa coupling matrices.

In the Next-to-Minimal Supersymmetric Standard Model (NMSSM), a
discrete $\mathbb{Z}_3$ symmetry is imposed, which forbids the tadpole and mass
terms and
only leaves dimensionless couplings in the superpotential. However, once the
singlet scalar acquires a vacuum expectation value, the $\mathbb{Z}_3$
symmetry is broken and unacceptable domain walls can be generated 
\cite{xmssm,domain}.

In the Nearly Minimal Supersymmetric Standard Model (nMSSM)
\cite{nMSSM,tadloop,hcorr,nMSSM2}, a $\mathbb{Z}_5$ or $\mathbb{Z}_7$
R-symmetry  is introduced in the superpotential and the K\"ahler potential,
which forbids  quadratic and cubic self-interactions of the singlet. However,
at higher loop orders, a tadpole term is induced due to supergravity effects.
It is suppressed at the six ($\mathbb{Z}_5$) or seven ($\mathbb{Z}_7$) loop
level, so that it is naturally of weak-scale order and does not destabilize the
hierarchy \cite{tadloop,hcorr}. The discrete symmetries also ensure that rapid
proton decay does not occur, while Majorana neutrino masses are allowed.
Furthermore the lightest supersymmetric particle, usually the lightest
neutralino, is quasi-stable on cosmological time-scales \cite{gaga}, even
without the introduction of R-Parity. The superpotential of the nMSSM is thus
\begin{equation}
\begin{aligned}
W_{\rm nMSSM} &= \lambda \hat{S} \hat{H}_1 \cdot \hat{H}_2 + 
  \frac{m^2_{12}}{\lambda}  \hat{S} \\
   & \;\;\;\; + y_{\rm u} \hat{Q} \cdot \hat{H}_2 \, \hat{U}^c 
         + y_{\rm d} \hat{Q} \cdot \hat{H}_1 \, \hat{D}^c
         + y_{\rm e} \hat{L} \cdot \hat{H}_1 \, \hat{E}^c.
\end{aligned}
\end{equation}
In the following, the phenomenological properties of the nMSSM will be analyzed
in more detail.

\subsection{CP violation}
\label{sc:cp}

In the nMSSM superpotential, the parameters $\lambda$, $m_{12}$ and $y_{\rm
f}$ can in general be complex. However, the phase of $\lambda$ does
not generate physical CP-violating effects, since it can be made real by
suitable gauge rotations \cite{realsup}. As in the Standard Model, the Yukawa
couplings $y_{\rm f}$ lead to one physical phase in the CKM quark mixing matrix,
which is constrained by present data from many
heavy-flavor experiments. The phase of $m_{12}^2$ will be addressed below.

Beyond the superpotential, new complex phases can appear through
supersymmetry breaking. The soft supersymmetry breaking Lagrangian reads
\begin{equation}
\begin{aligned}
{\cal L}_{\rm soft} &= m_1^2 H_1^\dagger H_1 + m_2^2 H_2^\dagger H_2 
 + \ms^2 |S|^2 + (\tad S + \al S H_1 \cdot H_2 + \mbox{h.c.}) \\
&\quad + (M_1 \widetilde{B}\widetilde{B} + 
 M_2 \widetilde{W} \cdot \widetilde{W} +
 M_3 \, \tilde{g}\tilde{g} + \mbox{h.c.}) \\
&\quad + m_{\tilde{q}}^2 \, \tilde{q}_\LL^\dagger \cdot \tilde{q}_\LL^{}
 + m_{\tilde{u}}^2 |\tilde{u}_\RR|^2 + m_{\tilde{d}}^2 |\tilde{d}_\RR|^2
 + m_{\tilde{l}}^2 \, \tilde{l}_\LL^\dagger \cdot \tilde{l}_\LL^{}
 + m_{\tilde{e}}^2 |\tilde{e}_\RR|^2 \\
&\quad + (y_{\rm u} A_{\rm u} \,\tilde{q}_\LL \cdot H_2
 \, \tilde{u}_\RR^* + y_{\rm d} A_{\rm d} \,\tilde{q}_\LL \cdot H_1
 \, \tilde{d}_\RR^* + y_{\rm e} A_{\rm e} \,\tilde{l}_\LL \cdot H_1
 \, \tilde{e}_\RR^* + \mbox{h.c.}).
\end{aligned}
\end{equation}
Here
$H_i,S,\tilde{q}_\LL,\tilde{u}_\RR,\tilde{d}_\RR,\tilde{l}_\LL,\tilde{e}_\RR$
are the scalar components of the superfields
$\hat{H}_i,\hat{S},\hat{Q},\hat{U},\hat{D},\hat{L},\hat{E}$, where the quark
and lepton fields exist in three generations (the generation index has been
suppressed in the formula). $\widetilde{B},\widetilde{W},\tilde{g}$ denote the
fermionic components of the gauge super-multiplets. Among the soft breaking
parameters, $a_\lambda$, $\tad$, $M_{1,2,3}$ and $A_{\rm u,d,e}$ can be complex.
However not all their phases are physical. To see this, one can observe that
the superpotential is invariant under an U(1)$_\RR$ symmetry, with the charges
listed in Tab.~\ref{tab:RPQ}. In addition, it obeys an approximate Peccei-Quinn
symmetry U(1)$_\text{PQ}$, which is broken by the singlet tadpole term
$\propto m^2_{12}$. Both U(1)$_\RR$ and U(1)$_\text{PQ}$ are broken by some of
the supersymmetry breaking terms.
%%%%%%%%%%%%%%%%%%%%%%%%%%%%%%%%%%%%%%%%%%%%%%%%%%%%%%%%%%%%%%%%%%%%
\renewcommand{\arraystretch}{1.2}
\begin{table}[tp]
\begin{center}
\begin{tabular}{|c|rrrrrrrrrrr|c|}
\hline
 & $\hat{H}_1$ & $\hat{H}_2$ & $\hat{S}$ & $\hat{Q}$ & $\hat{L}$ & $\hat{U}^c$ &
$\hat{D}^c$ & $\hat{E}^c$ & $\hat{B}$ & $\hat{W}$ & $\hat{g}$ & $W_{\rm nMSSM}$
\\
\hline
U(1)$_\RR$ & 0 & 0 & 2 & 1 & 1 & 1 & 1 & 1 & 0 & 0 & 0 & 2 \\
U(1)$_\text{PQ}$ & 1 & 1 & -2 & -1 & -1 & 0 & 0 & 0 & 0 & 0 & 0 & 0 \\
\hline
\end{tabular}
\end{center}
\vspace{-1em}
\mycaption{Charges of superfields under the U(1)$_\RR$ and U(1)$_\text{PQ}$
symmetries of the superpotential. Note that the U(1)$_\RR$ charges of the
fermionic components differ by 1 from those of the superfields.}
\label{tab:RPQ}
\end{table}
%%%%%%%%%%%%%%%%%%%%%%%%%%%%%%%%%%%%%%%%%%%%%%%%%%%%%%%%%%%%%%%%%%%%

With the help of the U(1)$_\RR$ and U(1)$_\text{PQ}$, the fields can be rotated
so that the phases of two parameters become zero. By analyzing the charges, it
can be seen that the following products remain invariant under both R- and
PQ-transformations:
\begin{equation}
\begin{aligned}
&\arg(m_{12}^2 \tad^* a_\lambda), \\
&\arg(m_{12}^2 \tad^* M_i), &\quad & i=1,2,3, \\
&\arg(m_{12}^2 \tad^* A_{\rm u}) , && \text{(3 generations)},\\
&\arg(m_{12}^2 \tad^* A_{\rm d}) , && \text{(3 generations)},\\ 
&\arg(m_{12}^2 \tad^* A_{\rm e}) , && \text{(3 generations)},
\end{aligned}
\end{equation}
corresponding to 13 physical CP-violating phases in addition to the CKM phase.
Without loss of generality, the phases of $m_{12}$ and $\tad$ can be
chosen real, so that the physical phases are transferred into $a_\lambda$,
$M_{1,2,3}$ and $A_{\rm u,d,e}$.

In this work, for
simplicity, gaugino unification is assumed, so that $M_1 : M_2 : M_3 \approx 1 :
2 : 6$. In this case, the gaugino masses carry one common phase, $\phi_{\rm M_1}
= \phi_{\rm M_2} =\phi_{\rm M_3} \equiv \phi_{\rm M}$. To simplify the analysis
further, the phases in $A_{\rm u,d,e}$ and $a_\lambda$ are set to zero.

\subsection{Higgs sector}

The scalar potential at tree-level receives contributions from F-, D-terms and
soft supersymmetry breaking. To avoid charge and color breaking vacuum
solutions and unacceptably large neutrino masses, 
the squarks and sleptons must not receive vacuum expectation values.
In this case, the Higgs potential at tree-level reads
\begin{align}
V_0 &= V_{\rm F}+V_{\rm D}+V_{\rm soft}, \\
V_{\rm F} &= 
 \left| \lambda H_1 \cdot H_2 + \frac{m_{12}^2}{\lambda} \right|^2 + 
 |\lambda S|^2 (H_1^\dagger H_1 + H_2^\dagger H_2),\\
V_{\rm D} &= \frac{\bar{g}^2}{8} (H_1^\dagger H_1 - H_2^\dagger H_2)^2 +
  \frac{g^2}{2} |H_1^\dagger H_2|^2, \\
V_{\rm soft} &= m_1^2 H_1^\dagger H_1 + m_2^2 H_2^\dagger H_2 
 + \ms^2 |S|^2 + (\tad S + \al S H_1 \cdot H_2 + \mbox{h.c.}),
\end{align}
where $\bar{g}^2 = g^2 + g'^2$, and $g$, $g'$ are the SU(2), U(1) gauge couplings,
respectively.
As mentioned in the previous section, we constrain all parameters in the Higgs
sector to be real.
Nevertheless, we allow complex phases in the gaugino sector, which in turn would generate 
small CP-violating phases in the Higgs sector through radiative corrections. 
However, the effect of gaugino loop contributions to the Higgs effective potential is 
generally sub-dominant, compared to the top-stop loops, and thus can be neglected in the 
following discussion.

In the zero-temperature vacuum state, the neutral Higgs components acquire
non-zero vacuum expectation values, 
\begin{equation}
\langle S \rangle = \vs, \qquad
\langle H_1^0 \rangle = v_1, \qquad
\langle H_2^0 \rangle = v_2.
\end{equation}
It is useful to define $\tan\beta = v_2/v_1$.
The minimization conditions for electroweak
symmetry breaking give at Born level
\begin{align}
m_1^2 &= -(m_{12}^2 + \al \vs) \frac{v_2}{v_1} - \frac{\bar{g}^2}{4}(v_1^2-v_2^2)
 - \lambda^2 (v_2^2 + \vs^2), \\
m_2^2 &= -(m_{12}^2 + \al \vs) \frac{v_1}{v_2} - \frac{\bar{g}^2}{4}(v_2^2-v_1^2)
 - \lambda^2 (v_1^2 + \vs^2), \\
\ms^2 &= -\al \frac{v_1 v_2}{\vs} - \frac{\tad}{\vs} - \lambda^2 v^2.
\end{align}
The physical mass eigenstates of the Higgs sector consist of three neutral
CP-even scalars $S_{1,2,3}$, two neutral CP-odd scalars $P_{1,2}$ and one
charged scalar $H^\pm$. Both the CP-even and CP-odd neutral scalars share a
component of the singlet scalar $S$.

In the sector of the CP-odd scalars, one combination of the two doublets form
the Goldstone mode $G^0$, which is absorbed into the longitudinal mode of the
$Z$ boson. The other linear combination of the two doublets, $A^0$, mixes with
the imaginary part of the singlet $S$ to give the two physical states. In the
basis $(A^0, \im S)$, their mass matrix reads
\begin{equation}
M_{\rm P}^2 = \begin{pmatrix}
\MA^2 & -\al v \\
-\al v & \ms^2 + \lambda^2 v^2
\end{pmatrix},
\end{equation}
where
\begin{equation}
\MA^2 = -\frac{2}{s_{2\beta}} (m_{12}^2 + \al\vs),
\end{equation}
and $s_{2\beta} = \sin 2 \beta$.
The mass eigenstates $P_{1,2}$ are related to this basis by the mixing matrix
$O^P$,
\begin{equation}
\begin{pmatrix}
P_1 \\ P_2 
\end{pmatrix} = O^P 
\begin{pmatrix}
A^0 \\ \im S 
\end{pmatrix}.
\end{equation}
For large values of $\MA$, corresponding to large negative values of
$m_{12}^2$, the mixing between the two CP-odd scalars is relatively small, with
the light pseudo-scalar $P_1$ being almost completely singlet, while the heavy
pseudo-scalar $P_2$ has a very small singlet component and a mass of about
$m_{\rm P_2} \sim \MA$. These properties have strong consequences for Higgs
searches at LHC and ILC, as discussed in the next section.

Similarly to the CP-odd sector, the CP-even Higgs eigenstates are related to the
doublet and singlet components by the mixing matrix $O^S$,
\begin{equation}
\begin{pmatrix}
S_1 \\ S_2 \\ S_3
\end{pmatrix} = O^S 
\begin{pmatrix}
H_1^0 \\H_2^0 \\ \re S 
\end{pmatrix}.
\end{equation}
For $\MA \ll \MZ$, there are one heavy mass eigenstate $S_3$ with small singlet 
component, and two light eigenstates $S_1$, $S_2$ with sizable singlet and doublet 
admixtures.

The Higgs potential receives large radiative corrections, with the dominant
contribution stemming from top-stop loops. These effects have been calculated
in Ref.~\cite{nmssmcorr,hcorr} and are included in this work.

The influence of the parameter $\lambda$ gives an additional positive
contribution to the mass of the lightest Higgs boson, so that the bound of
about 114 GeV from LEP can be easily avoided even for moderate values of
$\tan\beta$ and stop mass eigenvalues $\mst{1}, \mst{2}$. 
From the vacuum expectation value of $
S$, $\vs$, an effective $\mu$-parameter is generated, $\mu = -\lambda \vs$.

\subsection{Chargino and neutralino sector}

The presence of the fermion component of the singlet superfield, the singlino
$\tilde{S}$, leads to a fifth neutralino state. 
The chargino mass matrix in the basis $(\widetilde{W}^\pm, \widetilde{H}^\pm)$
is
\begin{equation}
M_{\cha^\pm} = \begin{pmatrix}
M_2 & \sqrt{2} s_\beta \MW \\
\sqrt{2} c_\beta \MW & -\lambda \vs 
\end{pmatrix},
\end{equation}
while the neutralino mass matrix in the basis $(\widetilde{B}^0,\widetilde{W}^0,
\widetilde{H}^0_1,\widetilde{H}^0_2,\widetilde{S})$
is given by
\begin{equation}
M_{\neu} = \begin{pmatrix}
M_1 & 0 & -c_\beta\sw\MZ & s_\beta\sw\MZ & 0 \\
0 & M_2 & c_\beta\cw\MZ & -s_\beta\cw\MZ & 0\\
-c_\beta\sw\MZ & c_\beta\cw\MZ & 0 & \lambda \vs & \lambda v_2 \\
s_\beta\sw\MZ & -s_\beta\cw\MZ & \lambda \vs & 0 & \lambda v_1 \\
0 & 0 & \lambda v_2 & \lambda v_1 & 0 \\
\end{pmatrix},
\end{equation}
where the abbreviations 
$s_\beta \equiv \sin\beta$, 
$c_\beta \equiv \cos\beta$, 
$\sw \equiv \sin\theta_{\rm W}$, 
$\cw \equiv \cos\theta_{\rm W}$ 
have been used.
The neutralino mass matrix is diagonalized by a unitary matrix $N$, such that
\begin{equation}
N^* M_{\neu} N^\dagger ={\rm diag}
(\mneu{1},\mneu{2},\mneu{3},\mneu{4},\mneu{5}). 
\end{equation}
Perturbativity of the trilinear coupling $\lambda$ up to the GUT scale
requires that $\lambda \lesim 0.8$ \cite{gaga}. 
For values of $\lambda < 0.8$, the lightest neutralino has a large singlino
component, and a mass below about 60 GeV.

\subsection{Experimental and astrophysical constraints}

The chargino and neutralino spectrum is constrained by new physics searches at
LEP. For the lightest chargino, the LEP measurements lead to the constraint
$\mcha{1} > 104$~GeV or $\sigma[e^+e^- \to \cha_1^+\cha_1^-] \times \text{BR} < 10$~fb, 
where $\sigma$ is the production cross section at $\sqrt{s} = 208$~GeV, and BR the 
branching ratio into hadron or leptons.
Similarly,
the corresponding requirement for neutralinos is $(\mneu{1}+\mneu{2}) > 208$~GeV or 
$\sigma[e^+e^- \to \neu_1\neu_2] \times \text{BR} < 10$~fb \cite{lepsusy}. 
Moreover,
if the lightest neutralino is so light that $\mneu{1} < \MZ/2$, the
measurement of the total $Z$-boson width imposes the limit BR$[Z\to
\neu_1\neu_1] < 0.8\times 10^{-3}$ \cite{lepz}. On the other hand, if the
nMSSM is to account for the dark matter relic density in the universe, 
the lightest neutralino mass is bound to be in the range 
$25 \gev < \mneu{1} < 40 \gev$~\cite{gaga}.

For the mechanism of electroweak baryogenesis to work successfully, the
electroweak phase transition needs to be strongly first order, and new sources
of CP-violation beyond the CKM matrix must be present. The electroweak phase
transition can be made first order through a relatively large cubic term in the
effective Higgs potential. In the nMSSM, similar to other singlet extensions of
the MSSM, this cubic term is realized at tree-level by the $a_\lambda$ soft
breaking parameter. As a consequence, a strongly first order phase transition
can be achieved without relying on radiative contributions from a light stop 
as in the MSSM \cite{gaga, huberco, NMSSMewbg}. As a result, neither of the stop
masses is constrained to be small ($\lesim 150 \gev$ in the MSSM), but rather can 
amount to several hundred GeV.

CP-violation in baryon-number violating processes can be introduced via
complex couplings in the Higgs sector or through complex parameters in currents
that couple sufficiently strongly to the Higgs potential. In the nMSSM, four
physical phases can thus contribute to the baryon asymmetry, the phases of
$M_1$, $M_2$, $A_{\rm t}$ and $\tad$ \cite{huberco}. In this work, as emphasized before,
it is assumed that the only non-zero phase (besides the CKM matrix) is a common
phase of the gaugino masses, $\phi_{\rm M_1}
= \phi_{\rm M_2} =\phi_{\rm M_3} \equiv \phi_{\rm M}$.

In the presence of CP-violating phases, the nMSSM parameter space is 
constrained by the experimental limits on the electric dipole moment (EDM) 
of the electron, neutron and $^{199}$Hg nucleus 
\cite{Regan:2002ta,Baker:2006}.  The upper bound on the electron EDM is 
derived from limits of the EDM of the $^{205}$Tl atom.  For the phases 
considered in this work, without mixing between the CP-even and CP-odd Higgs
states, the CP-odd 
electron-neutron operator studied in \cite{Pilaftsis:2002fe} vanishes, and 
the $^{205}$Tl EDM is due almost entirely to the electron EDM.  This 
translates into a limit on the electron EDM of \cite{Regan:2002ta} 
\begin{equation} 
|d_e| < 1.9\times 10^{-27}\;e\,cm, 
\label{Eq:de}
\end{equation} 
at 95\%~CL.  

The electron EDM receives potentially large contributions from one-loop 
sfermion-gaugino diagrams \cite{edm1}, which become small for large masses  of
several TeV for the first two generation sleptons.  A similar role is played by
the  squarks in the EDMs of the neutron and Hg atom.   Two-loop contributions
from diagrams with charginos and  charged Higgs propagators can also be
sizable, but just as in the MSSM  \cite{Abel:2001vy} they can be suppressed
below the current limits for  sizable values of the pseudo-scalar masses and
$\tan\beta \sim {\cal O}(1)$. However, since the two-loop contributions are enhanced by
the same complex phase that  generates the baryon asymmetry, the electron EDM
bound presents a  particularly severe constraint on electroweak baryogenesis.

The one- and two-loop contributions to the electron EDM were calculated 
following Refs. \cite{Abel:2001vy} and \cite{Chang:2002ex,Pilaftsis:2002fe}, 
respectively.  In the calculation of the one- and two-loop electron EDM 
contributions, we have included the Higgs and slepton mixing.  

Note that while slepton mixing is suppressed with the very small electron mass,
the overall contribution to the electron EDM requires a spin flip and is thus also 
proportional the electron mass. As a result, slepton mixing
can lead to non-negligible effects when the slepton masses are moderate.  On 
the other hand, the contribution from the heaviest of the Higgs 
states $S_3$ and $P_2$ is suppressed due to their large masses.

For first and second generation 
sfermion masses of about 10 TeV, pseudo-scalar mass $\MA$ higher than 500
GeV, and values of $\tan\beta \sim {\cal O}(1)$, which is in accordance with
baryogenesis constraints, we find that phases $\phi_{\rm M} \sim {\cal O}(1)$
are allowed by the electron and neutron EDM constraints. Since a sufficiently
large baryon asymmetry could be generated for phases in the range $\phi_{\rm M}
\sim 0.1 \dots 0.3$ \cite{gaga}% 
\footnote{Different methods have been used to compute the baryon asymmetry in
the MSSM~\cite{X,XX,XXX,XXXX}.
The values derived in Ref.~\cite{gaga} are based on the method of
Ref.~\cite{X}. While other methods~\cite{XXXX}
tend to lead to a higher baryon asymmetry, a more
recent calculation leads to
a lower one~\cite{XXX}. Furthermore, there are several
approximations performed in the computation of the baryon
asymmetry, leading to an uncertainty, of order one, on the necessary
CP-violation
phases required for the realization of this scenario. In section
\ref{sc:comment}, we provide a more general
discussion of the possible sources of CP-violation, and how they
affect the phenomenology of this model.}, 
this leaves a large window for this scenario to be
realized in the nMSSM.

Here we do not attempt to provide a theoretical model that can explain this pattern 
for the superpartner masses and parameters, but it has been pointed out in the literature
that a roughly similar pattern might emerge within the framework of split 
supersymmetry~\cite{splitnmssm}.

\subsection{Benchmark scenario A}

In the following sections, the collider phenomenology and dark matter detection
expectation will be analyzed in detail for the concrete parameter point A from
Ref.~\cite{gaga}. The parameters of this scenario are summarized in
Tab.~\ref{tab:A}.%
%%%%%%%%%%%%%%%%%%%%%%%%%%%%%%%%%%%%%%%%%%%%%%%%%%%%%%%%%%%%%%%%%%%%
\renewcommand{\arraystretch}{1.2}
\begin{table}[tp]
\begin{center}
\begin{tabular}{|ccccccccccc|}
\hline
$\tan\beta$ & $\lambda$ & $\vs$ & $\al$ & $\tad^{1/3}$ & $\MA$ & $|M_1|$ &
$|M_2|$ & $|M_3|$ & $\phi_{\rm M}$ & \\
\hline
1.7 & 0.619 & $-$384 & 373 & 157 & 923 & 122.5 & 245 & 730 & 0.14 & \\
\hline\hline
$m_{Q1,2}$ & $m_{U1,2}$ & $m_{D1,2}$ & $m_{L1,2}$ & $m_{E1,2}$ & 
$m_{Q3}$ & $m_{U3}$ & $m_{D3}$ & $m_{L3}$ & $m_{E3}$ & $A_{\rm t,b,\tau}$ \\
\hline
10000 & 10000 & 10000 & 10000 & 10000 & 
500 & 500 & 500 & 500 & 500 & $-$100 \\
\hline
\end{tabular}
\end{center}
\vspace{-1em}
\mycaption{Parameters for the reference point A \cite{gaga} used in this study.
All dimensionful parameters are given in GeV.}
\label{tab:A}
\end{table}
%%%%%%%%%%%%%%%%%%%%%%%%%%%%%%%%%%%%%%%%%%%%%%%%%%%%%%%%%%%%%%%%%%%%
To avoid EDM constraints, the sleptons and squarks of 
the first two generations have masses of 10 TeV in benchmark scenario A.
In contrast, the masses of the third generation sfermions are assumed to be
around 500 GeV. There are no strong bounds for these masses, but 
stop masses with a few hundred GeV are favored by Higgs mass naturalness
and baryogenesis.
The masses and decay modes of the neutralinos, charginos, Higgs scalars, third
generation squarks and the gluino for this scenario are listed in
Tab.~\ref{tab:decays}, Tab.~\ref{tab:decayh} and Tab.~\ref{tab:decsq}.
%%%%%%%%%%%%%%%%%%%%%%%%%%%%%%%%%%%%%%%%%%%%%%%%%%%%%%%%%%%%%%%%%%%%
\renewcommand{\arraystretch}{1.2}
\begin{table}[tp]
\begin{center}
\begin{tabular}{|c||c|c|r@{\:}ll|}
\hline
Sparticle & Mass $m$ [GeV] & Width $\Gamma$ [GeV]
          & \multicolumn{3}{c|}{Decay modes} \\
\hline \hline
$\neu_1$ & $33.3$ & --- & \multicolumn{2}{c}{---} & \\
$\neu_2$ & $106.6$  & $0.00004$ &
        $\neu_2$ & $\to Z^* \, \neu_1$ & 100\% \\
$\neu_3$ & $181.5$  & $0.09$ &
        $\neu_3$ & $\to Z \, \neu_1$ & 74\% \\
        &&&& $\to S_1 \, \neu_1$ & 26\% \\
        &&&& $\to P_1 \, \neu_1$ & 0.4\% \\
$\neu_4$ & $278.0$  & $1.5$ &
        $\neu_4$ & $\to Z \, \neu_1$ & 11\% \\
        &&&& $\to Z \, \neu_2$ & 22\% \\
        &&&& $\to Z \, \neu_3$ & 1\% \\
        &&&& $\to W^\pm \, \cha^\mp_1$ & 43\% \\
        &&&& $\to S_1 \, \neu_1$ & 7\% \\
        &&&& $\to S_1 \, \neu_2$ & 0.2\% \\
        &&&& $\to S_2 \, \neu_1$ & 8\% \\
        &&&& $\to P_1 \, \neu_1$ & 7\% \\
        &&&& $\to P_1 \, \neu_2$ & 0.7\% \\
$\neu_5$ & $324.4$  & $2.1$ &
        $\neu_5$ & $\to Z \, \neu_1$ & 30\% \\
        &&&& $\to Z \, \neu_2$ & 1.5\% \\
        &&&& $\to Z \, \neu_3$ & 0.15\% \\
        &&&& $\to W^\pm \, \cha^\mp_1$ & 57\% \\
        &&&& $\to S_1 \, \neu_1$ & 0.01\% \\
        &&&& $\to S_1 \, \neu_2$ & 0.02\% \\
        &&&& $\to S_1 \, \neu_3$ & 5\% \\
        &&&& $\to S_2 \, \neu_1$ & 1\% \\
        &&&& $\to S_2 \, \neu_2$ & 4\% \\
        &&&& $\to P_1 \, \neu_1$ & 0.4\% \\
        &&&& $\to P_1 \, \neu_2$ & 0.7\% \\
        &&&& $\to P_1 \, \neu_3$ & 0.06\% \\
\hline
$\cha_1^\pm$ & $165.0$  & $0.136$ &
        $\cha^+_1$ & $\to W^+ \, \neu_1$ & 100\% \\
$\cha_2^\pm$ & $319.5$  & $2.0$ &
        $\cha^+_2$ & $\to W^+ \, \neu_1$ & 32\% \\
        &&&& $\to W^+ \, \neu_2$ & 1\% \\
        &&&& $\to W^+ \, \neu_3$ & 34\% \\
        &&&& $\to Z \, \cha^+_1$ & 29\% \\
        &&&& $\to S_1 \, \cha^+_1$ & 5\% \\
        &&&& $\to P_1 \, \cha^+_1$ & 0.3\% \\
\hline
\end{tabular}
\end{center}
\vspace{-1em}
\mycaption{Masses, widths and main branching ratios of the
neutralino and chargino states at Born level
for the reference point A (Tab.~\ref{tab:A}).}
\label{tab:decays}
\end{table}
%%%%%%%%%%%%%%%%%%%%%%%%%%%%%%%%%%%%%%%%%%%%%%%%%%%%%%%%%%%%%%%%%%%%
\renewcommand{\arraystretch}{1.2}
\begin{table}[tp]
\begin{center}
\begin{tabular}{|c||c|c|r@{\:}lr@{}l|}
\hline
Sparticle & Mass $m$ [GeV] & Width $\Gamma$ [GeV]
          & \multicolumn{4}{c|}{Decay modes} \\
\hline \hline
$S_1$ & $115.2$ & $0.044$ & 
	$S_1$ & $\to b\bar{b}$ & 8&\% \\
	&&&& $\to \neu_1 \neu_1$ & 92&\% \\
$S_2$ & $156.6$ & $0.060$ & 
	$S_2$ & $\to b\bar{b}$ & 2&\% \\
	&&&& $\to  W^+W^-$ & 17&\% \\
	&&&& $\to  ZZ$ & 2&.5\% \\
	&&&& $\to \neu_1 \neu_1$ & 69&\% \\
	&&&& $\to \neu_1 \neu_2$ & 10&\% \\
\hline
$P_1$ & $133.7$ & $0.008$ & 
	$P_1$ & $\to \neu_1 \neu_1$ & 100&\% \\
\hline
\end{tabular}
\end{center}
\vspace{-1em}
\mycaption{Masses, widths and main branching ratios of the
light Higgs states, including one-loop top-stop corrections,
for the reference point A (Tab.~\ref{tab:A}).}
\label{tab:decayh}
\end{table}
%%%%%%%%%%%%%%%%%%%%%%%%%%%%%%%%%%%%%%%%%%%%%%%%%%%%%%%%%%%%%%%%%%%%
\renewcommand{\arraystretch}{1.2}

\begin{table}[tp]
\begin{center}
\begin{tabular}{|c||c|c|r@{\:}ll|}
\hline
Sparticle & Mass $m$ [GeV] & Width $\Gamma$ [GeV]
          & \multicolumn{3}{c|}{Decay modes} \\
\hline \hline
$\st_1$ & $522.1$  & $43.1$ &
        $\st_1$ & $\to t \neu_1$ & 12\% \\
        &&&& $\to t \neu_2$ & 7\% \\
        &&&& $\to t \neu_3$ & 11\% \\
        &&&& $\to t \neu_4$ & 20\% \\
        &&&& $\to t \neu_5$ & 1\% \\
        &&&& $\to b \cha^+_1$ & 47\% \\
        &&&& $\to b \cha^+_2$ & 2\% \\
$\st_2$ & $535.3$  & $29.8$ &
        $\st_2$ & $\to t \neu_1$ & 14\% \\
        &&&& $\to t \neu_2$ & 2\% \\
        &&&& $\to t \neu_3$ & 11\% \\
        &&&& $\to t \neu_4$ & 9\% \\
        &&&& $\to t \neu_5$ & 20\% \\
        &&&& $\to b \cha^+_1$ & 6\% \\
        &&&& $\to b \cha^+_2$ & 37\% \\
\hline
$\sB_1$ & $498.8$  & $6.6$ &
        $\sB_1$ & $\to b \neu_1$ & 0.5\% \\
        &&&& $\to b \neu_2$ & 12\% \\
        &&&& $\to b \neu_3$ & 12\% \\
        &&&& $\to b \neu_4$ & 0.1\% \\
        &&&& $\to b \neu_5$ & 5\% \\
        &&&& $\to t \cha^-_1$ & 57\% \\
        &&&& $\to t \cha^-_2$ & 15\% \\
$\sB_2$ & $503.3$  & $12.5$ &
        $\sB_2$ & $\to b \neu_1$ & 0.6\% \\
        &&&& $\to b \neu_2$ & 4\% \\
        &&&& $\to b \neu_3$ & 9\% \\
        &&&& $\to b \neu_4$ & 0.2\% \\
        &&&& $\to b \neu_5$ & 7\% \\
        &&&& $\to t \cha^-_1$ & 56\% \\
        &&&& $\to t \cha^-_2$ & 23\% \\
\hline
$\glo$ & $730$ & 6.6 & $\glo$ & $\to b\sB_1$ & 35\% \\
	&&&& $\to b\sB_2$ & 34\% \\
	&&&& $\to t\st_1$ & 18\% \\
	&&&& $\to t\st_2$ & 13\% \\
\hline
\end{tabular}
\end{center}
\vspace{-1em}
\mycaption{Masses, widths and main branching ratios of the
third generation squarks and the gluino at Born level
for the reference point A (Tab.~\ref{tab:A}).}
\label{tab:decsq}
\end{table}
%%%%%%%%%%%%%%%%%%%%%%%%%%%%%%%%%%%%%%%%%%%%%%%%%%%%%%%%%%%%%%%%%%%%
The relatively light neutralino and chargino spectrum is typical for a nMSSM
scenario that is in agreement with electroweak baryogenesis and the relic
dark matter density \cite{gaga}. The lightest neutralino has a large singlino
component, so
that the bound from the invisible $Z$-boson width is evaded although
$\mneu{1} < \MZ/2$. Nevertheless, the dark matter annihilation cross-section is
dominated by s-channel $Z$ exchange, since the combined mass of two $\neu_1$ is
close to the $Z$ resonance, $2\mneu{1} \sim \MZ$.
As the sfermions are relatively heavy, the other neutralino and chargino
states mainly decay through gauge bosons.

Owing the small mass of the lightest neutralino $\neu_1$, the light Higgs
scalars $S_1$, $S_2$ and $P_1$ decay predominantly into two $\neu_1$.
For the light CP-odd scalar $P_1$, which is an almost pure singlino state, this
is essentially the only allowed channel, with all other contributions being far
below 1\% and thus negligible.

Since the most relevant contribution to the annihilation cross section is 
coming from the $Z$ pole, in Ref. \cite{gaga} a simple calculation sufficed 
to determine the relic abundance of neutralinos for the benchmark 
cases.\footnote{We thank A.~Menon and D.~Morrissey to provide us with 
their original codes.}  Due to the precision nature that the ILC lends to 
our present work, we paid careful attention to reproducing, and  
improving, the earlier results.  In our work, we include an improved 
superpartner spectrum calculation, all possible co-annihilation channels, 
all SM final states, and solve the relevant Boltzmann equation numerically 
for the relic density calculation.

While our numbers closely agree with those of Ref. \cite{gaga} there are  some
small deviations.  Notably, our relic density calculation results in  slightly
higher values.  When using the input values for benchmark A, we  obtain a
neutralino relic density of $\Omega h^2 = 0.131$.  The deviation of  this from
the latest WMAP central value $\Omega h^2 = 0.111$ \cite{Spergel:2006hy} is
comparable  to the other theoretical and experimental uncertainties (higher
order  corrections to sparticle masses and annihilation cross sections,
systematic  errors, etc.) entering into this calculation.  

To demonstrate the uncertainty in the relic density calculation, we note 
that a variation of the neutralino mass by less than 1 GeV is enough to 
change the relic density by 0.025 for scenario A.  Since the main focus of 
our work is collider phenomenology and a 1 GeV shift in the neutralino mass 
affects collider phenomenology negligibly, our results remain valid for 
all values close to $\Omega h^2 = 0.111$.

Therefore, in spite of the discrepancy in $\Omega h^2$, we will stick to the scenario A in our numerical analysis.

%%%%%%%%%%%%%%%%%%%%%%%%%%%%%%%%%%%%%%%%%%%%%%%%%%%%%%%%%%%%%%

\section{Collider measurements in the nMSSM}
\label{sc:coll}

The nMSSM can play a crucial role in baryogenesis and dark matter generation,
depending on the parameters of the Higgs, chargino and neutralino sectors. In
this section it is studied how the relevant particles can be discovered and the
relevant parameters be determined at future colliders.

\subsection{The nMSSM at LHC}

\subsubsection{Higgs physics}

In the benchmark scenario A, the light Higgs states $S_1$, $S_2$ and $P_1$
mainly decay invisibly into the lightest neutralinos.
This behavior is rather typical for many singlet-extended supersymmetry models \cite{Lang07}.
A CP-even invisible Higgs boson can be discovered at the LHC through $W$-boson
fusion \cite{wbf} and through associated $Zh$ production \cite{zh}.
The cross-sections and final
state distributions for these processes in the nMSSM are the same as for the
Standard Model Higgs boson, except for modified Higgs-$Z$ couplings, that are
given by
\begin{equation}
G_{\rm ZZS_i} = G^{\rm SM}_{\rm ZZH} (s_\beta O^S_{i1} + c_\beta O^S_{i2}).
\label{eq:zzs}
\end{equation}
For the reference point A (Tab.~\ref{tab:A}), the couplings for the three
CP-even Higgs states amount to
\begin{align}
G_{\rm ZZS_1} &= 0.87 \times G^{\rm SM}_{\rm ZZH},\nonumber\\
G_{\rm ZZS_2} &= 0.49 \times G^{\rm SM}_{\rm ZZH},\label{eq:zzsn}\\
G_{\rm ZZS_3} &= 0.002 \times G^{\rm SM}_{\rm ZZH}.\nonumber
\end{align}
Since both the $S_1$ and $S_2$ states have ${\cal O}(1)$ couplings to the $Z$ boson and are 
relatively light, with masses below 200~GeV, they can be produced with sizable rates.
Based on the analysis of Ref.~\cite{wbf}, it can be estimated that with only a
few
fb$^{-1}$ of integrated luminosity, a 5$\sigma$ discovery in the $W$-boson
fusion channel can be achieved in the scenario A.

For the case of a single Standard-Model-like Higgs boson, the mass of the Higgs
can be determined from the ratio of the $W$-boson
fusion and $Zh$ production rates \cite{dhl}. On the other hand, in scenario A,
the invisible Higgs signal receives contributions from both the $S_1$ and $S_2$.
However, the observables in both
$W$-boson
fusion and $Zh$ production are not sensitive to discriminate between one and two
invisible Higgs states. As a result, from an invisible Higgs signal at the LHC,
it is not possible to obtain information about the number and the masses of the
CP-even Higgs bosons in the nMSSM.

\subsubsection{Supersymmetric particles} 

Since the partners of quarks and 
the gluon, squarks and gluino, couple with the strong QCD coupling, they are 
produced with large cross-sections at the LHC. Charginos and neutralino can 
be generated in the decay cascades of squarks and the gluino with sizeable 
rates. In principle, charginos and neutralino are also produced directly in 
electroweak processes, but the cross-sections for these channels are small. 
Therefore in the following only squarks and gluinos are considered as 
primary supersymmetric particles. At the Tevatron, the typically large 
slepton masses necessary to suppress the electron electric dipole moment in 
this model lead to a reduced branching ratio for the decay of neutralinos 
and charginos into lepton final states. This makes their searches in the 
tri-lepton channel quite difficult, particularly for masses of the chargino 
and second lightest neutralino larger than 150~GeV, as are typical in the 
nMSSM scenario under analysis~\cite{Beate}.

The relevant leading-order production cross-sections for squarks and gluinos at 
the LHC are summarized in Tab.~\ref{tab:lhcxsec}, calculated 
with {\sc CompHEP 4.4} \cite{comphep}.
%%%%%%%%%%%%%%%%%%%%%%%%%%%%%%%%%%%%%%%%%%%%%%%%%%%%%%%%%%%%%%%%%%%%
\renewcommand{\arraystretch}{1.2}
\begin{table}[tp]
\begin{center}
\begin{tabular}{|c|ccccccccc|}
\hline
$pp \to$ & $\glo\glo$ & 
 $\sB_1\sB^*_1$ & $\sB_1\sB^*_2$/$\sB_2\sB^*_1$ & 
 $\sB_2\sB^*_2$ &
 $\st_1\st^*_1$ & $\st_1\st^*_2$/$\st_2\st^*_1$ & 
 $\st_2\st^*_2$ &
 $\glo\sB^{(*)}_1$ & $\glo\sB^{(*)}_2$ \\
\hline \hline
$\sigma$ [fb] & 2162 & 444 & 3 & 421 & 357 & 1 & 312 & 141 & 138 \\
\hline
\end{tabular}
\end{center}
\vspace{-1em}
\mycaption{Tree-level production cross-sections for strongly interacting 
supersymmetric particles at the LHC
for the reference point A, see Tab.~\ref{tab:A}. The QCD scale is taken at the average
mass of the produced particles, $Q = (m_{\tilde{x}} + m_{\tilde{y}})/2$.}
\label{tab:lhcxsec}
\end{table}
%%%%%%%%%%%%%%%%%%%%%%%%%%%%%%%%%%%%%%%%%%%%%%%%%%%%%%%%%%%%%%%%%%%%
The analysis of $\neu_2$ production is experimentally particularly promising
\cite{lhclc}. The neutralino $\neu_2$ is produced in various squark and gluino
decay cascades, leading to a total cross-section for $\neu_2$ production with
leptonic $\neu_2$ decays of 30 fb. 
Here the most important channel is
\begin{equation}
pp \to \tilde{g}\tilde{g}, \qquad
\tilde{g} \to 
b \tilde{b}^{*} \text{ or } \bar{b} \tilde{b}
%\raisebox{.25ex}{$\stackrel{\mbox{\tiny (--)}}{b}$} 
%\tilde{b}^{(*)} 
\to b \bar{b} \neu_2,
\end{equation}
but direct production of sbottoms and stops via $pp \to \tilde{b}\tilde{b}^*,
\tilde{t}\tilde{t}^*$ also plays a role.
According to Ref.~\cite{mill,lhclc}, 
background from SM gauge bosons can be reduced by
cuts on missing transverse energy and missing mass:
\begin{itemize}
\item At least three jets with transverse momentum $p_t^{\rm jet} > 150,100,50$
GeV.
\item Missing energy $\Eslash > \max(100 \gev, 0.2 M_{\rm eff})$ with
$M_{\rm eff} \equiv \Eslash + \sum_{i=1}^3 p_{t,i}^{\rm jet}$.
\item Two isolated leptons with $p_t^{\rm lep} > 20,10$ GeV.
\end{itemize}
The remaining $t\bar{t}$ background is removed by 
subtracting events with two different-flavor leptons from
events with same-flavor leptons. This procedure makes use of the fact that the
$t\bar{t}$ background produces the same number of same-flavor and different-flavor
lepton pairs, while the 
neutralino signal has only same-flavor lepton pairs. After these cuts
practically no SM background is left, while a signal efficiency for $\neu_2$
production of better than 20\% is achieved
\cite{mill,lhclc}. This corresponds to about 1800 signal events for an integrated 
luminosity of 300 fb$^{-1}$.

The two-lepton signal for $\neu_2$ production can also originate from the
neutralino $\neu_3$, whereas the contamination from heavier neutralinos is very
small. The total cross-section for leptonic $\neu_3$ decays is 40 fb.
Contrary to the $\neu_2$, the two leptons from $\neu_3$ originate from a real
$Z$-boson and have an invariant mass equal to $\MZ$.

\begin{figure}[tb]
\centering
\epsfig{figure=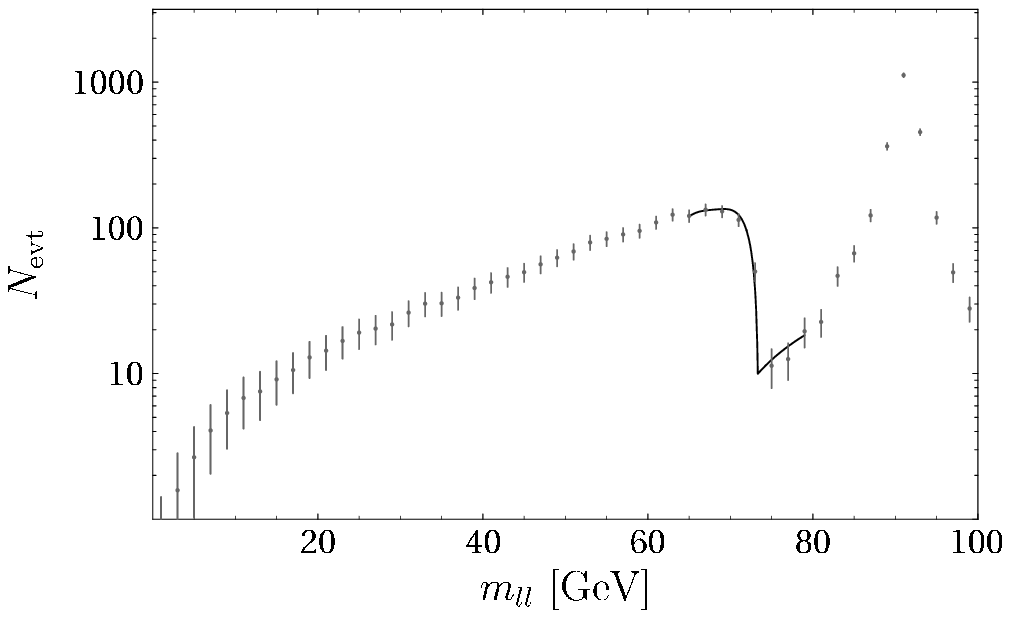, width=11cm, bb=0 505 304 678}
\mycaption{Fit to $m_{ll}$ distribution for light neutralino
production at the LHC. Backgrounds from Standard Model sources are not included,
as they are expected to be small.}
\label{fg:lhc}
\end{figure}

For the scenario A, see Tab.~\ref{tab:A}, the production of neutralinos $\neu_2$
and $\neu_3$ has been simulated with {\sc CompHEP 4.4} \cite{comphep},
using CTEQ6M parton distribution functions.
The production cross-section is substantially modified by QCD corrections
\cite{nlolhc}. However, for the determination of superpartner masses, only the
kinematic properties of the decay products are important, which are modified
relatively little by radiative corrections. For the purpose of this work,
radiative corrections have thus been neglected.
Information about superpartner masses can be extracted from 
kinematic edges in invariant mass
spectra of the final state particles \cite{lhclc,lhcedge}.
The distribution of the di-lepton invariant mass $m_{ll}$ in $\neu_2$ decay has a 
sharp upper edge 
\begin{equation}
m_{ll,\rm max,2} = \mneu{2} - \mneu{1},
\end{equation}
see Fig.~\ref{fg:lhc}.
%%%%%%%%%%%%%%%%%%%%%%%%%%%%%%%%%%%%%%%%%%%%%%%%%%%%%%%%%%%%%%
%%%%%%%%%%%%%%%%%%%%%%%%%%%%%%%%%%%%%%%%%%%%%%%%%%%%%%%%%%%%%%
The peak at $m_{ll} = \MZ$ comes from the contribution of $\neu_3$, while events
at lower invariant masses originate mainly from the $\neu_2$.
Assuming 300 fb$^{-1}$ luminosity,
a simple fit to the upper edge of that region gives 
\begin{equation}
m_{ll,\rm max,2} = 73.5 \pm 0.5 \pm 0.08 \gev,	\label{eq:ll}
\end{equation}
where the first error is statistical, while the second error accounts for 
the systematic error from energy
scale uncertainty in the detector (see \cite{mill} for discussion).
The error is comparable to what was found in \cite{mill} for the MSSM scenario
($\beta$).

For further studies, the decay chains involving the $\neu_3$ can be separated from the 
$\neu_2$ by applying the cut $|m_{ll} - \MZ| < 10$~GeV on the di-lepton invariant mass.
Including the jet from the squark decay $\sB \to b \neu_i$ gives additional
information. For the decay chain with the $\neu_3$ , the invariant $m_{jll,3}$ 
distribution has an upper endpoint with 
\begin{equation}
\begin{aligned}
m_{jll,\rm max,3}^2 = \frac{1}{2\mneu{3}^2}
 \bigl[ \mneu{1}^2\mneu{3}^2 - \mneu{3}^4 - \mneu{1}^2 m_{\sB}^2 + 
 \mneu{3}^2 m_{\sB}^2 + \mneu{3}^2\MZ^2 + m_{\sB}^2\MZ^2 \\ -
  (\mneu{3}^2 - m_{\sB}^2) 
  \sqrt{\lambda(\mneu{1}^2,\mneu{3}^2,\MZ^2)} \bigr].
\end{aligned}
\end{equation}
with $\lambda(a,b,c) = a^2 + b^2 + c^2 - 2 a b - 2 a c - 2 b c$.
Since the mass difference between $m_{\sB_1}$ and $m_{\sB_2}$ is small, no
experimental distinction between the two states can and needs to be made.
Flavor-tagging of the b-jet from the sbottom decay does not improve the analysis,
since the main background is $t\bar{t}$.

%%%%%%%%%%%%%%%%%%%%%%%%%%%%%%%%%%%%%%%%%%%%%%%%%%%%%%%%%%%%%%
\begin{figure}[tb] 
\anc \hspace{1ex} \makebox[0mm][l]{(a)} \hspace{8cm} (b) \\[1.5ex]
\epsfig{figure=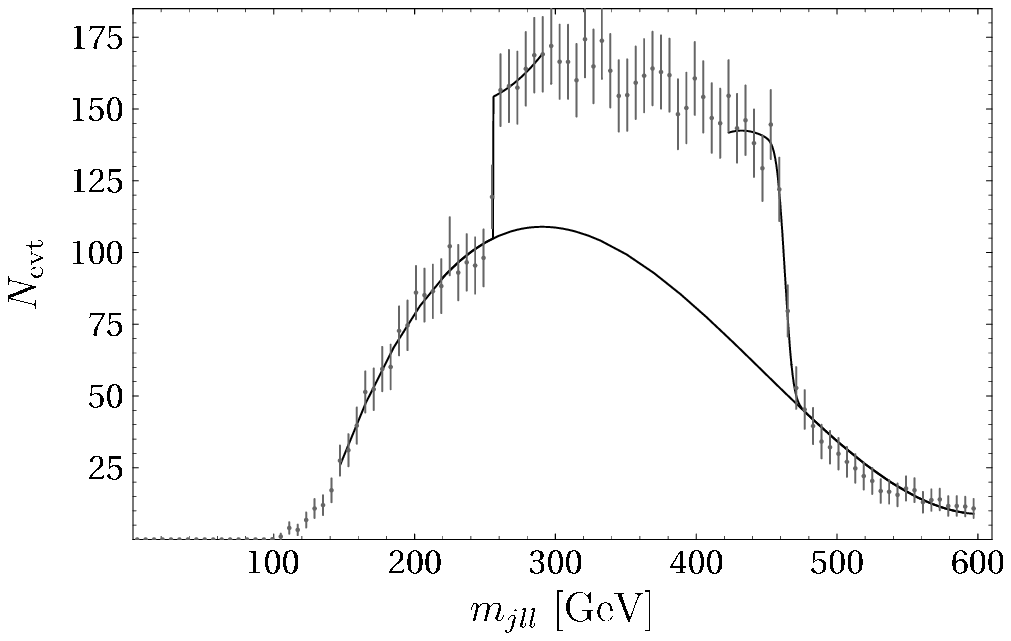, width=8cm, bb=20 505 300 682} \hspace{1ex}
\epsfig{figure=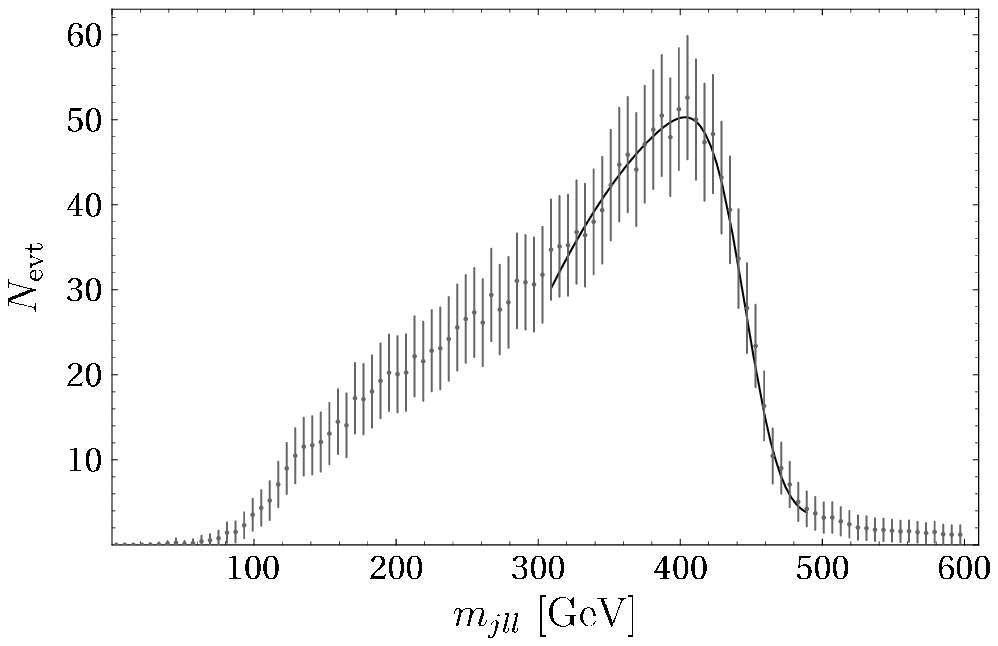, width=8cm, bb=20 505 300 682}
\mycaption{Fits to the $m_{jll}$ distribution for (a) $\neu_3$ and (b) $\neu_2$
production at the LHC. Backgrounds from Standard Model sources are not included,
as they are expected to be small.}
\label{fg:lhc2}
\end{figure}
%%%%%%%%%%%%%%%%%%%%%%%%%%%%%%%%%%%%%%%%%%%%%%%%%%%%%%%%%%%%%%
In a typical supersymmetry event, there are multiple jets. The jet from $\sB \to b \neu_i$ is
expected to be relatively hard $E_{\rm T,j} \gesim 200$ GeV, but there are
additional hard jets from the decay of the other sbottom and from gluinos, $\glo \to
b \sB$. This introduces an irreducible combinatorial background. However, including
that background, the characteristic edge in the $m_{jll,3}$ distribution at
$m_{jll,\rm max,3}$ is still visible, see Fig.~\ref{fg:lhc2}. The combinatorial
background can be reduced by special techniques \cite{mill,lhclc}, 
but here we simply choose to fit it. The fit result is
\begin{equation}
m_{jll,\rm max,3} = 463.6^{+5.5}_{-9.0} \pm 2.3 \gev,\label{eq:jll3a}
\end{equation}
where as before the second error includes lepton and jet energy scale
uncertainties.
A second edge in the $m_{jll,3}$ distribution is found at
\begin{equation}
\begin{aligned}
m_{jll,\rm min,3}^2 = \frac{1}{2\mneu{3}^2}
 \bigl[ \mneu{1}^2\mneu{3}^2 - \mneu{3}^4 - \mneu{1}^2 m_{\sB}^2 + 
 \mneu{3}^2 m_{\sB}^2 + \mneu{3}^2\MZ^2 + m_{\sB}^2\MZ^2 \\ +
  (\mneu{3}^2 - m_{\sB}^2) 
  \sqrt{\lambda(\mneu{1}^2,\mneu{3}^2,\MZ^2)} \bigr],
\end{aligned}
\end{equation}
which can be fitted in the same way as the upper end point, yielding
\begin{equation}
m_{jll,\rm min,3} = 256.2^{+6.0}_{-7.0} \pm 1.3 \gev. \label{eq:jll3b}
\end{equation}
In addition to studying the decay chain with the $\neu_3$,
by requiring the invariant mass of the lepton pair to be sufficiently below the
$Z$ pole, $m_{ll} < \MZ - 10 \gev$, the decay chain with the $\neu_2$ can be
selected. Similarly to the $\neu_3$ case, 
the $m_{jll,\rm max,2}$ distribution has a characteristic endpoint at 
\begin{equation}
m_{jll,\rm max,2}^2 = \frac{1}{\mneu{2}^2} (\mneu{2}^2 - \mneu{1}^2)
(m_{\sB}^2 - \mneu{2}^2).
\end{equation}
As the $\neu_2$ decays through an off-shell $Z^*$, the $m_{jll,\rm max,2}$
distribution has no characteristic endpoint towards the lower end. To first
approximation, the spectrum of $\neu_2$ decays via an off-shell $Z^*$ can be
thought of as superposition of Breit-Wigner line-shapes, which are close to
Gaussian. Consequently, the upper end of the $m_{jll,\rm max,2}$ distribution 
can be approximated by an error function. A fit gives the rather poor result
\begin{equation}
m_{jll,\rm max,2} = 447^{+14}_{-21}\pm 2.3 \gev, \label{eq:jll2}
\end{equation}
which is limited by statistics and the shape of the distribution near
the endpoint, which is less steep than for the di-lepton distribution.

Light charginos $\cha^\pm_1$ can be detected in the squark decay chains by
looking for a same-sign lepton signal originating from the processes
\begin{equation}
\begin{aligned}
pp &\to \tilde{g}\tilde{g} \to bb\tilde{b}^*\tilde{b}^* \to 
bb\,\bar{t}\bar{t}\, \cha^+_1 \cha^+_1 \to bb\,\bar{t}\bar{t}\, W^+W^+ \,
\neu_1\neu_1
\to bb\,\bar{t}\bar{t}\, l^+ l^+ \,\nu_l \nu_l\, \neu_1\neu_1, \\
pp &\to \tilde{g}\tilde{g} \to \bar{b}\bar{b}\,\tilde{b}\tilde{b} \to 
\bar{b}\bar{b}\,tt \,\cha^-_1 \cha^-_1 \to \bar{b}\bar{b}\,tt\, W^-W^-\, \neu_1\neu_1
\to \bar{b}\bar{b}\,tt\, l^- l^-\, \bar{\nu}_l \bar{\nu}_l\, \neu_1\neu_1, \\
pp &\to \tilde{g}\tilde{g} \to tt\tilde{t}^*\tilde{t}^* \to 
tt\,\bar{b}\bar{b}\, \cha^-_1 \cha^-_1 \to tt\,\bar{b}\bar{b}\, W^-W^- \,
\neu_1\neu_1
\to tt\,\bar{b}\bar{b}\, l^- l^-\, \bar{\nu}_l \bar{\nu}_l\, \neu_1\neu_1, \\
pp &\to \tilde{g}\tilde{g} \to \bar{t}\bar{t}\,\tilde{t}\tilde{t} \to 
\bar{t}\bar{t}\,bb \,\cha^+_1 \cha^+_1 \to \bar{t}\bar{t}\,bb\, W^+W^+\, \neu_1\neu_1
\to \bar{t}\bar{t}\,bb\, l^+ l^+ \,\nu_l \nu_l\, \neu_1\neu_1, \\
\end{aligned}
\end{equation}
see Ref.~\cite{sqcd}. However, since besides the
neutralino as the lightest supersymmetric particle, the neutrino in the
chargino decay also escapes detection, the remaining lepton-jet invariant mass
distributions do not allow a meaningful mass extraction.

The measurement of the heavy neutralinos $\neu_4$ and $\neu_5$ at the LHC is
very difficult. As pointed out above, the appearance of a lepton pair in the
neutralino decay is the best possibility for detection. However, due to small
branching ratios of the heavy neutralinos into leptons, the statistics for this
channel are very low.

\begin{figure}[tb]
\centering
\epsfig{figure=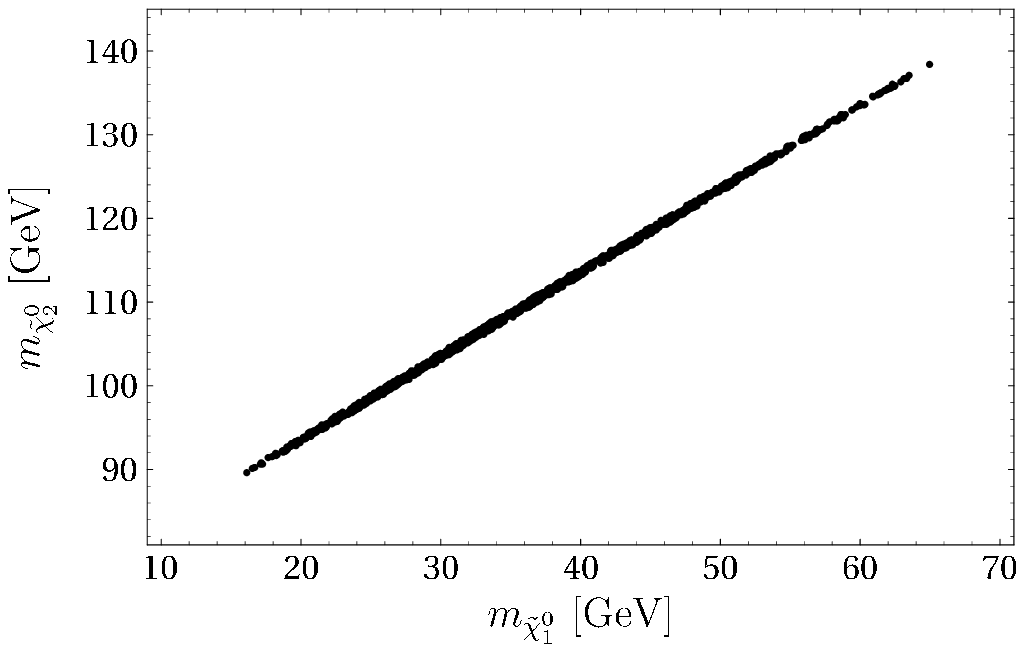, width=11cm, bb=0 505 304 681}\hspace{1cm}\ 
\mycaption{Correlation between $\mneu{1}$ and $\mneu{2}$ from LHC measurements.}
\label{fg:lhccorr}
\end{figure}

From the combination of the results in eqs.~\eqref{eq:ll}, \eqref{eq:jll3a},
\eqref{eq:jll3b}, and \eqref{eq:jll2} one can extract the following
absolute values for the superpartner masses,
\begin{align}
\mneu{1} &= 33^{+32}_{-17.5} \gev, & 
\mneu{2} &= 106.5^{+32.5}_{-17.5} \gev, &
\mneu{3} &= 181^{+20}_{-10} \gev, &
m_{\sB} &= 499^{+30}_{-17} \gev. \label{eq:lhcres}
\end{align}
The large errors are due to large correlations between the mass parameters, 
as illustrated for one example in Fig.~\ref{fg:lhccorr}.
%%%%%%%%%%%%%%%%%%%%%%%%%%%%%%%%%%%%%%%%%%%%%%%%%%%%%%%%%%%%%%
%%%%%%%%%%%%%%%%%%%%%%%%%%%%%%%%%%%%%%%%%%%%%%%%%%%%%%%%%%%%%%
This can be explained by the fact that all measurements of kinematic endpoints 
in the decay distributions are closely related to mass \emph{differences},
whereas no independent direct measurement of one of the masses, e.g. the
lightest neutralino mass, is available.

The analysis in this section has been performed for the specific parameter
point A (see Tab.~\ref{tab:A}). However, most of the results are expected to be
rather typical for nMSSM scenarios that explain baryogenesis and dark matter. 
These two constraints predict that all neutralino and chargino states should be
relatively light. In addition, our LHC analysis relies heavily on the gluino and a
sbottom state with large left-chiral component to have masses of the order of a
few hundred GeV. At present, 
the mass of the gluino is not constrained by any data, but if
GUT-scale gaugino unification is realized, the gluino should have a mass of
about 1 TeV or less. As discussed before, the presence of light squarks
are helpful in avoiding the suppression of chiral charges leading to the
generation of the baryon asymmetry in this model. Following
Ref.~\cite{gaga} we have therefore assumed
that the third generation squarks have masses of the order of 500~GeV. 
While, as emphasized above, this assumption is crucial in the analysis of the 
LHC phenomenology, it plays only a minor role in the ILC related one.
At an $e^+e^-$ collider such as the ILC, a precise measurement of the lightest
neutralino mass is possible, so that by combing results from the LHC and ILC, the
precision of the neutralino mass determination can be greatly improved. 
This was
discussed already for the MSSM in Ref.~\cite{lhclc} and will be 
analyzed for the
nMSSM in the next sections.

\subsection{The nMSSM at Giga-Z}

In nMSSM scenarios that account for baryogenesis and cold dark matter, the
lightest neutralino mass is typically smaller than half the $Z$-boson mass,
$\mneu{1} < \MZ/2$. In particular this is the case in the reference point A
(cf.~Tab.~\ref{tab:decays}). As a consequence, the lightest neutralino contributes
to the invisible $Z$ width. Current limits from LEP \cite{lepinv} already pose
strong constraints on the nMSSM parameter space \cite{lepz}. Nevertheless,
these constraints can be improved at a future high-luminosity linear collider
(ILC) running on the $Z$ pole \cite{gigaz}. This "Giga-Z" option for the ILC
could collect 50 fb$^{-1}$ of integrated luminosity at $\sqrt{s} \sim \MZ$,
which is an improvement of several orders of magnitude compared to LEP.

Since in a typical scenario like reference point A, the neutralino dark matter
annihilation proceeds almost exclusively via s-channel $Z$-exchange, the precise
determination of the invisible $Z$
width opens up the opportunity to directly and model-independently determine the
relevant $Z\neu_1\neu_1$ coupling 
\begin{equation}
|G_{Z\neu_1\neu_1}|^2 = \frac{g^2}{4\cw^2}\left(|N_{13}|^2 -
|N_{14}|^2\right)^2.
\label{GZn1n1}
\end{equation}
However, the achievable accuracy at Giga-Z for $\Gamma^{\rm Z}_{\rm inv}$ is
limited by systematics, and can reach at best 0.1\% \cite{invZ}, 
which is only a factor
4 better than current LEP constraints. 
This corresponds to an error of about $0.2 \times 10^{-3}$ for BR$[Z\to
\neu_1\neu_1]$.
In the case of scenario A, BR$[Z\to \neu_1\neu_1] \approx 0.3 \times 10^{-3}$,
so that the dark matter relic density can be determined from the invisible $Z$ width
with an uncertainty of at best 60\%.
Here it is assumed that the mass $\mneu{1}$ is determined from
some other observable with a much smaller error. 

Despite the large error, a Giga-Z
analysis can confirm the existence of a neutral, \mbox{(quasi-)stable}, 
weakly interacting particle and a rough
estimate of its coupling to the $Z$ boson in a model-independent way. This
would also be
an interesting cross-check for more detailed model-dependent measurements at
higher energies.

\subsection{The nMSSM at ILC}

\subsubsection{Higgs physics}

The CP-even Higgs bosons can be produced in $e^+e^-$ collisions through
radiation off a $Z$ boson, $e^+e^- \to S_i Z$. This process is very similar to
the case of the Standard Model Higgs boson, except for the different Higgs-$Z$
couplings strength as in eq.~\eqref{eq:zzs}. The $S_1$ and $S_2$ states have
relatively large couplings to the $Z$ boson, see eq.~\eqref{eq:zzsn}, 
and are therefore produced in sizeable rates.

As mentioned above,  both $S_1$ and $S_2$ decay predominantly invisibly into
the lightest neutralino in scenario A. Nevertheless, the kinematic mass peaks
of the Higgs
bosons can be reconstructed from the recoil of the $Z$, which is
cleanly characterized by the leptonic $Z$ decays $Z \to e^+e^-$ and $Z \to
\mu^+\mu^-$ \cite{hmass,schum}. Due to photon radiation and detector resolution
effects, the reconstructed mass peaks are smeared out somewhat, but
their width is less than 10 GeV, so that the two states $S_1$ and $S_2$ can be
clearly distinguished.

Taking into account the reduced $G_{\rm ZZS_i}$ couplings with respect to the
SM, the Higgs masses can be determined from the $Z$ recoil spectrum with the
precision \cite{hmass}
\begin{equation}
\delta M_{S1} \approx 130 \mev, \qquad
\delta M_{S2} \approx 185 \mev.
\end{equation}
Based on similar studies for the SM-like and invisible Higgses
\cite{SMhiggs,schum}, the branching ratios of $S_1$ and
$S_2$ can be extracted with the following accuracy,
\begin{gather}
\begin{aligned}
BR[S_1 \to b\bar{b}] &= (8 \pm 0.7)\%, & \qquad
 BR[S_1 \to {\rm inv.}] &= (91 \pm 3)\%, \\
BR[S_2 \to b\bar{b}] &= (2 \pm 0.3)\%, &
 BR[S_2 \to {\rm inv.}] &= (79 \pm 5)\%, 
\end{aligned} \\
BR[S_2 \to W^+W^-] = (17 \pm 1.5)\%. \nonumber
\end{gather}
Here only the statistical error is given, taking into account selection cuts to
reduce the backgrounds. The large invisible branching
ratio of both scalar states points towards a sizeable Higgs self-coupling
$\lambda$.

The light pseudo-scalar $P_1$ is an almost pure singlet, and the production
cross-section $e^+e^- \to S_i P_1$ suppressed by more than two orders of
magnitude with respect to the corresponding process in the MSSM. Furthermore,
the only available final state $b\bar{b} + \Eslash$ is swamped by background
from $S_i Z$ production, so that it appears impossible to discover the
$P_1$ at the ILC.

The two heavy neutral Higgs bosons $S_3$ and $P_2$ as well as the
charged Higgs boson are too heavy be produced at the ILC with center-of-mass
energies up to 1 TeV. In particular the charged Higgs boson should always have
a clearly visible decay channel into top- and bottom-quarks, $H^+ \to t
\bar{b}$. Charged Higgs production proceeds mainly through an s-channel
photon (with only a sub-dominant contribution from s-channel $Z$ exchange)
and thus cannot be suppressed by modified couplings.
As a consequence, the non-observation of the charged Higgs would clearly
indicate that $\MA$ is large, $\MA > \sqrt{s}/2$.
In this case, from the observation of the two light invisibly decaying Higgs
states, it can be deduced that these two Higgs bosons must be mixtures of the
SM-like Higgs and a new singlet. In other words, from the analysis of the Higgs
bosons at the 500 GeV ILC, one can already identify
an extended Higgs sector beyond the two doublets in the MSSM.

\subsubsection{Supersymmetric particles}

In scenario A, many of the neutralino
and chargino states are produced with sizeable cross-sections at the ILC with
$\sqrt{s} = 500$ GeV, as shown in Tab.~\ref{tab:ilcXsec}.
The most promising production processes are $e^+e^- \to \neu_2\neu_4, \;
\neu_3\neu_4, \; \cha^+_1\cha^-_1, \; \cha^\pm_1\cha^\mp_2$, with
cross-sections of more than 10 fb each. While the lightest neutralino is not
produced directly with large rates, it can nevertheless be studied in great
detail in the decays of the heavier neutralinos and charginos.
%%%%%%%%%%%%%%%%%%%%%%%%%%%%%%%%%%%%%%%%%%%%%%%%%%%%%%%%%%%%%%%%%%%%
\renewcommand{\arraystretch}{1.2}
\begin{table}[tp]
\begin{center}
\begin{tabular}{|r||rrrr|}
\hline
$e^+e^- \to \neu_i \neu_j$ & $\neu_i = \neu_2$ & $\neu_3$ & $\neu_4$ & $\neu_5$
\\
\hline \hline
$\neu_j = \neu_1$ & 2.0 & 5.4 & 3.7 & 3.9 \\
$\neu_2$ & 0.4 & 0.6 & 16.2 & 0.1 \\
$\neu_3$ && 0.1 & 32.8 & --- \\
$\neu_4$ &&& --- & --- \\
$\neu_5$ &&&& --- \\
\hline \hline
$e^+e^- \to \cha^\pm_i \cha^\mp_j$ & $\cha^\pm_i = \cha^\pm_1$ & $\cha^\pm_2$ &&
\\
\hline \hline
$\cha^\mp_j = \cha^\mp_1$ & 594 & 32.2 && \\
$\cha^\mp_2$ & & --- && \\
\hline
\end{tabular}
\end{center}
\vspace{-1em}
\mycaption{Tree-level production cross-sections in fb at $\sqrt{s} = 500$ GeV
with unpolarized beams
for the reference point A (Tab.~\ref{tab:A}).}
\label{tab:ilcXsec}
\end{table}
%%%%%%%%%%%%%%%%%%%%%%%%%%%%%%%%%%%%%%%%%%%%%%%%%%%%%%%%%%%%%%%%%%%%

In the following subsections, the most promising decay channels will be
analyzed in detail, including Standard Model and supersymmetric backgrounds. It
is found that the two charginos and all neutralinos except the $\neu_5$ could be
discovered and their masses and cross-sections measured. The discovery of a
fifth neutralino state would be a smoking gun for a non-minimal supersymmetric
model, but turns out to be very challenging experimentally. Nevertheless, the
discovery of two neutralino states which are much lighter than the lightest
chargino is already clear evidence for physics beyond the MSSM, where only one
neutralino, a dominant bino-state, can be significantly lighter than the
charginos.

We simulated signal and background with the Monte-Carlo methods from
\cite{slep}, including full tree-level matrix elements and Breit-Wigner
parameterizations for resonant intermediate particles. Initial-state radiation
and beamstrahlung are always included.
The processes are generated at the parton level, but jet
energy fluctuations through parton shower and detector effects are parameterized
by smearing functions with lepton and jet energy uncertainties taken from
\cite{tesladet}. Jets overlapping within a cone with $\Delta R =
\sqrt{(\phi_1-\phi_2)^2 + (\eta_1-\eta_2)^2} < 0.3$ are combined into one jet,
where $\phi_i$ and $\eta_i$ are the azimuthal angle and rapidity of jet $i$.
Similarly, a lepton lying within a jet is combined into the jet. Leptons and
jets outside the central region of the detector have a higher likelihood of
mistag and get inflicted by the large two-photon background. Therefore leptons
within an angle of $|\cos \theta| < 0.95$ around the beam line and jets with
$|\cos \theta| < 0.90$ are discarded. After these simple procedures, the
remaining isolated jets and leptons define the signature of the simulated
event.

For most processes, the signal cross-sections can be enhanced and background
can be reduced by a suitable choice of beam polarization. Here we assume that
both the electron and positron beam are polarized, with polarization degrees of
80\% and 50\%, respectively. It is further assumed that 500 fb$^{-1}$ of
luminosity is spent for $P(e^+)$/$P(e^-)$ = left/right and right/left each. The
center-of-mass energy is always $\sqrt{s} = 500$ GeV. The signal
and main SM background cross-sections for polarized beams are summarized in 
Tab.~\ref{tab:polXsec}.
%%%%%%%%%%%%%%%%%%%%%%%%%%%%%%%%%%%%%%%%%%%%%%%%%%%%%%%%%%%%%%%%%%%%
\renewcommand{\arraystretch}{1.2}
\begin{table}[tp]
\begin{center}
\begin{tabular}{|c||rrrr|rrrr|}
\hline
$P(e^+)$/$P(e^-)$ & $\neu_2\neu_4$ & $\neu_3\neu_4$ & $\cha^+_1\cha^-_1$ & 
$\cha^\pm_1\cha^\mp_2$ &
$W^+W^-$ & $ZZ$ & $t\bar{t}$ & $W^+W^-Z$\\
\hline \hline
80\% left / 50\% right & 25.8 & 52.2 & 1557 & 51.3 & 24500 & 1020 &
1130 & 95\phantom{.5} 
\\
80\% right / 50\% left & 19.6 & 39.7 & 107 & 39.0 & 770 & 440 & 500 & 4.5
\\
\hline
\end{tabular}
\end{center}
\vspace{-1em}
\mycaption{Polarized tree-level
production cross-sections in fb for neutralino, chargino
and some of the largest SM background processes at $\sqrt{s} = 500$ GeV
for the reference point A (Tab.~\ref{tab:A}).}
\label{tab:polXsec}
\end{table}
%%%%%%%%%%%%%%%%%%%%%%%%%%%%%%%%%%%%%%%%%%%%%%%%%%%%%%%%%%%%%%%%%%%%

For all neutralino and chargino processes under study here, the main SM
background come from double and triple gauge boson production, $t\bar{t}$
production and two-photon processes. Quite generally, they can be reduced by
observing that the supersymmetric signal processes lead to large missing
energy, and most of the final state particles go in the central detector
region. In particular, by imposing a minimum value for the total transverse
momentum, $p_t > 12$ GeV, the two-photon background is practically completely
removed\footnote{Note that the rejection of the two-photon and $e^\pm$-$\gamma$
background depends crucially on an excellent coverage of the detector at low
polar angles, so that energetic fermions with low transverse momentum can be
vetoed. The results of Ref.~\cite{stop} are based on the detector design of the
TESLA study \cite{tesladet}, with low beam crossing angle, muon detectors
extending to 65 mrad, and endcap calorimeters extending to 27.5 mrad. Although
for the current ILC detector R \& D several changes in the details of this
setup are discussed, the planned ILC detector designs are expected to reach a
similar photon-induced background rejection~\cite{GWilson}. However,  we also
want to point out that the  simulation of the photon-induced background in
Ref.~\cite{stop} with PYTHIA  \cite{pythia} has unquantified and possibly large
theoretical uncertainties.} \cite{stop}. Furthermore, cuts on the missing
energy $\Eslash$, the polar angle of the missing momentum, $\cos \theta_{p_{\rm
miss}}$, and of the visible momentum, $\cos \theta_{p_{\rm tot}} = p_{\rm
long}/p_{\rm tot}$ are effective to reduce the Standard Model backgrounds.

\subsubsection{Chargino \boldmath $\cha_1^+$}

The lightest charginos almost exclusively decay via
$\cha_1^\pm \to W^\pm \neu_1$, where the $W$ can decay leptonically or
hadronically. Since charginos are produced in pairs, two $W$ bosons appear in
their decays. If both $W$'s decay leptonically, the rates are relatively low
and one has to deal with a difficult background from $W^+W^-$. The purely
hadronic final state, on the other hand, is plagued by jet pair combination
ambiguities. Therefore the best mode is: $e^+e^- \to \cha_1^+ \cha_1^- \to W^+
W^- \,
\neu_1 \, \neu_1  \to j j \, l^\pm + \Eslash$ where $j$ stands for a jet.
The most important SM backgrounds are $W^+W^-$, $ZZ$, $t\bar{t}$ and two-photon
production, while supersymmetric backgrounds from neutralino pairs, $ZS_i$ and
Higgs pairs are also taken into account.

The SM backgrounds can be reduced by the general cuts explained in the previous
section, $p_t > 12$ GeV, $\Eslash > 100$ GeV and $|\cos \theta_{p_{\rm miss}}| <
0.8$. Since the two jets in the signal originate from a $W$ boson, the
invariant jet mass is required to fulfill $|m_{jj} - \MW| < 10$ GeV, which
removes neutralino background and
is also effective on $t\bar{t}$.
The $WW$ and $t\bar{t}$ backgrounds are further reduced by placing a cut on the
reconstructed invariant mass of lepton and missing momentum, $m_{l,\rm miss} >
150$ GeV.

Both the signal and the main background are increased for the beam polarization
combination $P(e^+)$/$P(e^-)$ = right/left, but since the signal cross-section
is large, and backgrounds after cuts are relatively low, this polarization
combination helps to increase the measurement precision.

With the selection cuts listed above, about $S= 105000$ signal events are
retained while only $B=30000$ background events survive (mainly from Standard
Model sources). The statistical error for total cross-section measurement is
$\delta\sigma^\pm_{11,\rm L} = 0.35$\%. For the opposite polarization
combination, $P(e^+)$/$P(e^-)$ = left/right, we obtain
$\delta\sigma^\pm_{11,\rm R} = 1.3$\%.

The distribution of the $W$-boson energy from chargino decay, reconstructed
from the momenta of the two jets, can be used for a chargino mass
measurement. The 
energy spectrum of the $W$ boson is almost evenly
distributed with characteristic endpoints at
\begin{equation}
\begin{aligned}
E_{\rm min,max} &= \frac{1}{4 \mcha{1}^2} \Bigl [ 
(\mcha{1}^2 - \mneu{1}^2 + \MW^2) \sqrt{s} \mp
  \sqrt{\lambda(\mcha{1}^2, \mneu{1}^2, \MW^2)\bigl(s - 4 \mcha{1}^2 \bigr)}
    \Bigr ] . \label{eq:chaend}
\end{aligned}
\end{equation}

\begin{figure}[tb]
\centering
\epsfig{figure=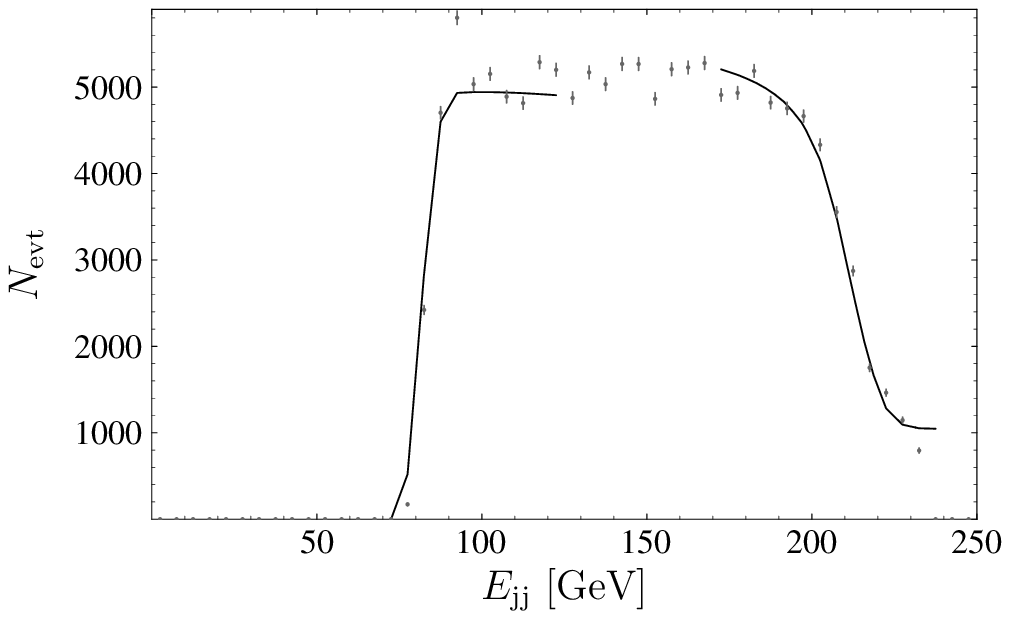, width=11cm}\hspace{1cm}\anc
\mycaption{Energy distribution for the jet pair originating from the $W$ boson
in the decay $\cha_1^\pm \to W^\pm \neu_1$ after selection cuts, with a simple
fit to the kinematic endpoints.}
\label{fg:chafit}
\end{figure}

The energy distribution edges can be fitted by using a step function which is
convoluted with the initial-state radiation and beamstrahlung spectrum and with
the jet smearing function. 
The fit function is fitted to the distribution obtained from the Monte-Carlo
simulation with a binned $\chi^2$ fit. 
The fit results are (see Fig.~\ref{fg:chafit})
%%%%%%%%%%%%%%%%%%%%%%%%%%%%%%%%%%%%%%%%%%%%%%%%%%%%%%%%%%%%%%
%%%%%%%%%%%%%%%%%%%%%%%%%%%%%%%%%%%%%%%%%%%%%%%%%%%%%%%%%%%%%%
\begin{align}
E_{\rm min} &= 83.73^{+0.025}_{-0.011} \gev & E_{\rm max} &= 214.8 \pm 0.8 \gev.
\label{eq:chafit}
\end{align}
resulting in
\begin{align}
\mcha{1} &= 165.0 \pm 0.3 \gev &  \mneu{1} &= 33.3^{+1.2}_{-1.1} \gev.
\end{align}
Note that only the statistical error has been given here, while the analysis of
systematic errors at this level of precision would require a more elaborate
investigation. Experience from the $W$ mass measurement at LEP however shows
that the systematic errors can be controlled to better than this level of
accuracy.

\subsubsection{Threshold scan for chargino \boldmath $\cha^\pm_1$}

In the determination of the chargino and lightest neutralino masses from the
decay distribution, the precision is limited due to substantial correlation
between $\mcha{1}$ and $\mneu{1}$ in the analysis. This can be improved by using
an independent measurement of $\mcha{1}$ via a threshold scan.

As an example, the measurement of the chargino pair production cross-section at
six different center-of-mass energies below and above the nominal threshold
$E_{\rm thr} = 2\mcha{1} = 330$ GeV are considered. Here it is assumed that the
chargino mass is already roughly
known from distribution measurements studied above.
The measurements below the
threshold allow to determine the background and extrapolate to values $\sqrt{s}
> 2\mcha{1}$, where the chargino excitation curve sets in. In combination with
measurements above the threshold, the threshold energy  $E_{\rm thr}$ can be
precisely determined. 

\begin{figure}[tb]
\centering
\epsfig{figure=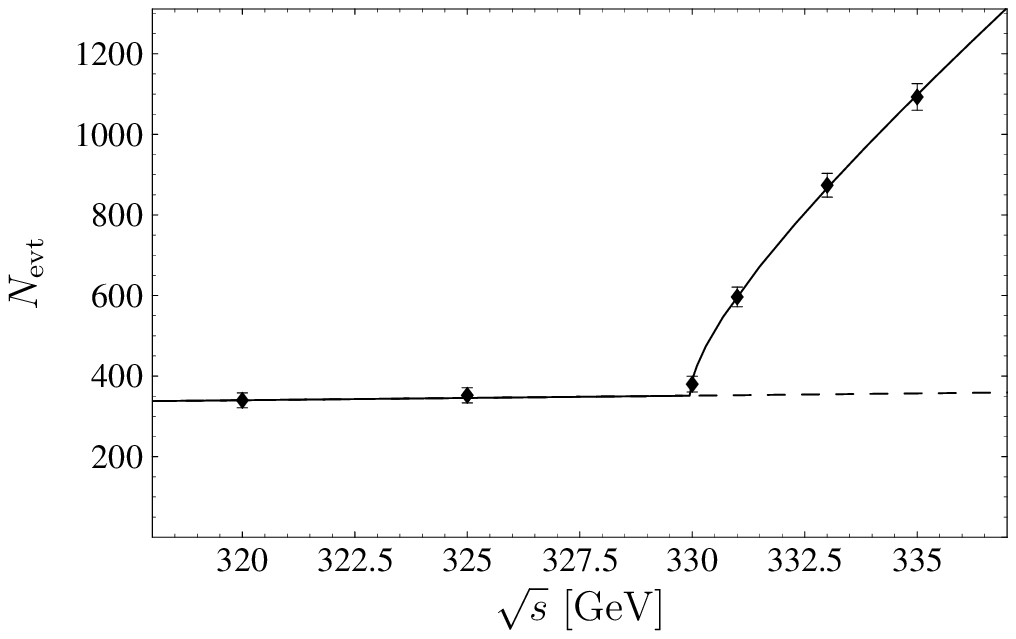, width=11cm}
\mycaption{Threshold scan for chargino pair production at ILC.}
\label{fg:thrscan}
\end{figure}

It is assumed that 10 fb$^{-1}$ luminosity is spent per point, amounting to
total of 60 fb$^{-1}$. As before, the beams are polarized with  $P(e^+)$/$P(e^-)$
= right/left, and the same selection cuts as in the previous subsection are
applied. Since the chargino mass can already be determined from decay
distributions, this information together with eq.~\eqref{eq:chaend} can be used
to reduce the background further.
The result of a simulation performed with this procedure is shown in
Fig.~\ref{fg:thrscan}.
%%%%%%%%%%%%%%%%%%%%%%%%%%%%%%%%%%%%%%%%%%%%%%%%%%%%%%%%%%%%%%
%%%%%%%%%%%%%%%%%%%%%%%%%%%%%%%%%%%%%%%%%%%%%%%%%%%%%%%%%%%%%% 
A fit with a simple quadratic function gives a very small statistical error,
\begin{equation}
E_{\rm thr} = 329.97 \pm 0.1 \pm 0.03 \gev,
\end{equation}
where the second error comes from the uncertainty in the incoming beam energy,
which is estimated to be of the order $10^{-4}$ \cite{icfa}. This corresponds to
$\mcha{1} = 164.98 \pm 0.05 \gev$. Together with the information from kinematic
edges in eq.~\eqref{eq:chafit}, the lightest neutralino mass is constrained to
\begin{equation}
\mneu{1} = 33.3^{+0.4}_{-0.3} \gev.
\end{equation}

\subsubsection{Neutralino \boldmath $\neu_3\neu_4$}

Both $\neu_3$ and $\neu_4$ have sizeable branching ratios into $Z$ bosons,
BR$[\neu_3 \to Z \neu_1] = 74$\% and BR$[\neu_4 \to Z \neu_1] = 11$\%.
This decay has the advantage of leading to very clear final state signatures,
and small background contamination from SM processes involving $W$-boson.
Because of small branching ratio of the $Z$ into
leptons, only the hadronic final states are promising for the
analysis of the neutralinos, at the cost of dealing
with higher backgrounds and jet pair combination
ambiguities.
Then the signature is: $e^+e^- \to \neu_3 \neu_4 \to ZZ \,
\neu_1 \, \neu_1  \to 4j + \Eslash$.

The largest SM backgrounds arise from $ZZ$, $W^+W^-$, $t\bar{t}$ and two-photon
processes, but supersymmetric backgrounds from production of other neutralino
pairs, $e^+e^- \to \neu_i \neu_j$, with $(i,j) \neq (3,4)$, and
chargino pairs, $e^+e^- \to \cha_1^+ \cha_1^-$, are also important.
The SM backgrounds can be suppressed by choosing $P(e^+)$/$P(e^-)$ =
left/right, while the neutralino signal remains sizeable for this
polarization combination. With the general cuts mentioned above, $p_t > 12$ GeV,
$\Eslash > 100$ GeV and $|p_{\rm long}/p_{\rm tot}| < 0.9$, the backgrounds are
further reduced. In addition, two pairs of jets have to form the invariant $Z$
mass,
$|m_{j_1j_2} - \MZ| < 10$ GeV and $|m_{j_3j_4} - \MZ| < 10$ GeV, which
removes $\neu_2 \neu_i$ background and
is also effective on $t\bar{t}$.
The $t\bar{t}$ background is reduced even further by using an
anti-bottom-tag with efficiency 95\% and a mistag rate 
of 3\% for light flavors and 25\% for charm jets \cite{btag}.
Finally, the chargino background is cut by removing events where 
two jets combine to give the invariant $W$ mass, $|m_{j_ij_j} - \MW| < 5$ GeV.

After these cuts, $S = 400$ signal events and only $B = 38$ background events
remain, leading to a statistical error for total cross-section measurement of
$\delta\sigma^0_{34} = 5.2$\%.

\begin{figure}[tb]
\centering
\epsfig{figure=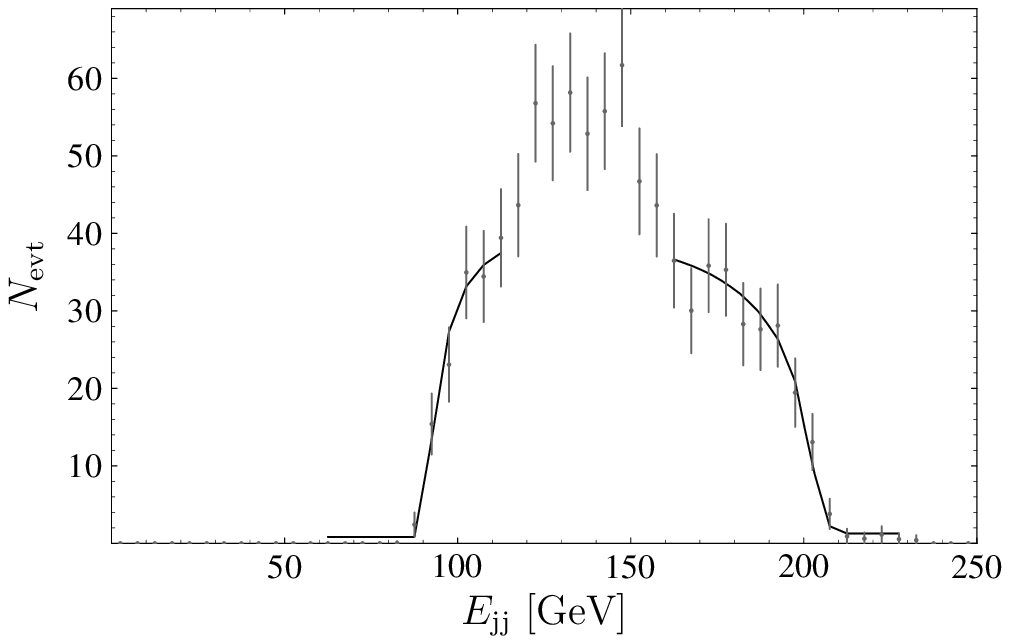, width=8cm}
\epsfig{figure=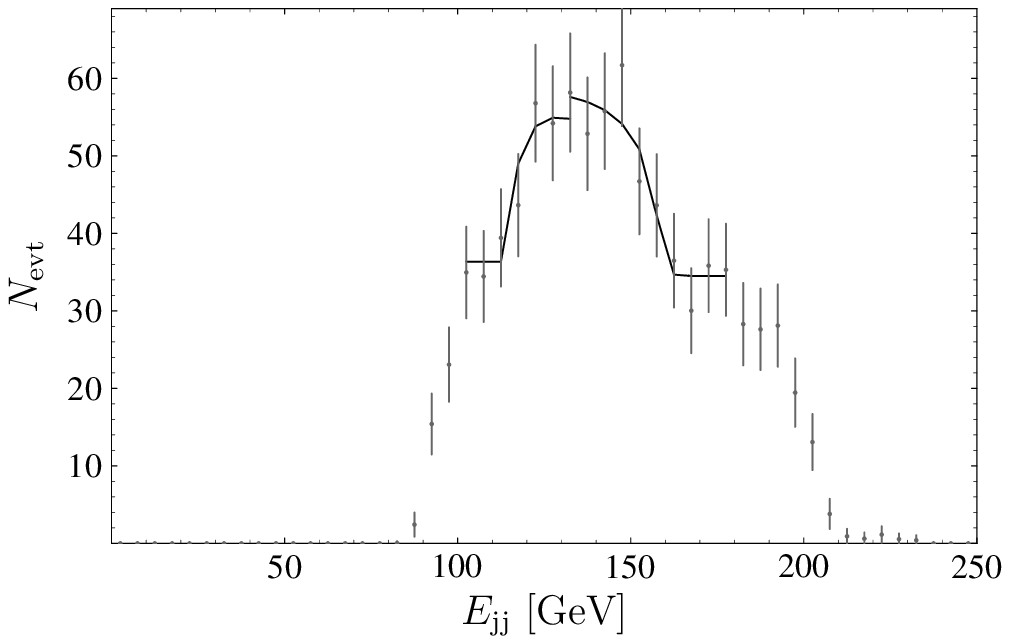, width=8cm}
\mycaption{Energy distribution for the jet pair originating from $Z$ bosons in
the decays of the neutralinos for the process $e^+e^- \to \neu_3 \neu_4$. The
plots show the expected event rates after selection cuts,  with a simple fit
to the kinematic edges.}
\label{fg:neu34fit}
\end{figure}

The energy spectrum of the $Z$-bosons from neutralino decay  has 
characteristic upper and lower endpoints both for the $\neu_3$ and $\neu_4$,
given by the expressions
\begin{equation}
\begin{aligned}
E_{\rm min,max,3} &= \frac{1}{4\mneu{3}^2\sqrt{s}} 
\Bigl ( \mneu{3}^4 - \mneu{3}^2\mneu{4}^2 + \mneu{3}^2\MZ^2 -
 \mneu{4}^2\MZ^2 + \mneu{3}^2 s + \MZ^2 s \\ & \qquad\qquad\qquad - \mneu{1}^2 (\mneu{3}^2 -
 \mneu{4}^2 + s) \mp \sqrt{\lambda(\mneu{3}^2, \mneu{1}^2, \MZ^2) \,
   \lambda(\mneu{3}^2, \mneu{4}^2, s)}\Bigr ),\\
E_{\rm min,max,4} &= \frac{1}{4\mneu{4}^2\sqrt{s}} 
\Bigl ( \mneu{4}^4 - \mneu{3}^2\mneu{4}^2 + \mneu{4}^2\MZ^2 -
 \mneu{3}^2\MZ^2 + \mneu{4}^2 s + \MZ^2 s \\ & \qquad\qquad\qquad - 
 \mneu{1}^2 (\mneu{4}^2 -
 \mneu{3}^2 + s) \mp \sqrt{\lambda(\mneu{4}^2, \mneu{1}^2, \MZ^2) \,
   \lambda(\mneu{3}^2, \mneu{4}^2, s)}\Bigr ),
\end{aligned}
\end{equation}
Since we consider the same decay mode for the $Z$ bosons stemming from both
neutralinos, a distinction between the decay products of the $\neu_3$ and
$\neu_4$ is not possible. Therefore, the measured spectrum contains all four
kinematic edges at the same time, 
which can be fitted with a convoluted step function as above.
From the fit, see Fig.~\ref{fg:neu34fit},
%%%%%%%%%%%%%%%%%%%%%%%%%%%%%%%%%%%%%%%%%%%%%%%%%%%%%%%%%%%%%%
%%%%%%%%%%%%%%%%%%%%%%%%%%%%%%%%%%%%%%%%%%%%%%%%%%%%%%%%%%%%%%
one obtains
\begin{align}
\mneu{3} &= 181.5 \pm 7.6 \gev &  \mneu{4} &= 278.0 \pm 11.5 \gev,
\end{align}
while no good constraint on $\mneu{1}$ is obtained. The masses of both $\neu_3$
and $\neu_4$ can be determined, but due to the low statistics, the error is
relatively large.

\subsubsection{Neutralino \boldmath $\neu_2\neu_4$}

The analysis of $\neu_2\neu_4$ production works similarly to the $\neu_3\neu_4$
production described in the previous subsection. The main difference lies in
the fact that the $\neu_2$, due to small mass difference to the lightest
neutralino $\neu_1$, decays only through a virtual, not on-shell, $Z$ boson.
Again, the best statistical significance of the signal is achieved by focusing
on the hadronic decay modes of the $Z$ bosons, $e^+e^- \to \neu_2 \neu_4 \to
ZZ^* \,
\neu_1 \, \neu_1  \to 4j + \Eslash$, and choosing the polarization combination
$P(e^+)$/$P(e^-)$ = left/right.

The backgrounds from $ZZ$, $W^+W^-$, $t\bar{t}$ and two-photon
processes as well as $\neu_3\neu_4$ neutralino and
$e^+e^- \to \cha_1^+ \cha_1^-$ chargino production can be further
reduced by the usual selection cuts, $p_t > 50$ GeV,
$\Eslash > 100$ GeV, $|p_{\rm long}/p_{\rm tot}| < 0.9$,
$|m_{j_ij_j} - \MW| < 5$ GeV, and an anti-b-tag.
Because of the large mass difference between the two produced neutralinos
$\neu_2$ and $\neu_4$, the signal typically has a relatively large total
transverse momentum, so that the cut on $p_t$ is increased to 50 GeV here.
Moreover, one pair of jets, stemming from the $\neu_4$ decay, has to have an
invariant mass equal to the $Z$-boson mass, while the other jet pair, associated
with the $ \neu_2$, must have an invariant mass smaller then the $Z$-boson mass.
Therefore the cuts $|m_{j_1j_2} - \MZ| < 10$ GeV and $\MZ - m_{j_3j_4} > 10$ GeV
are very effective.

These cuts reduce the background to $B
= 61$ events, while $S = 430$ signal events are retained. Therefore the total
production cross-section can be extracted with a statistical error of 
$\delta\sigma^0_{24} = 5.4$\%.

\begin{figure}[tb]
\begin{tabular}{p{8.1cm}p{7.9cm}}
(a) & (b) \\
\epsfig{figure=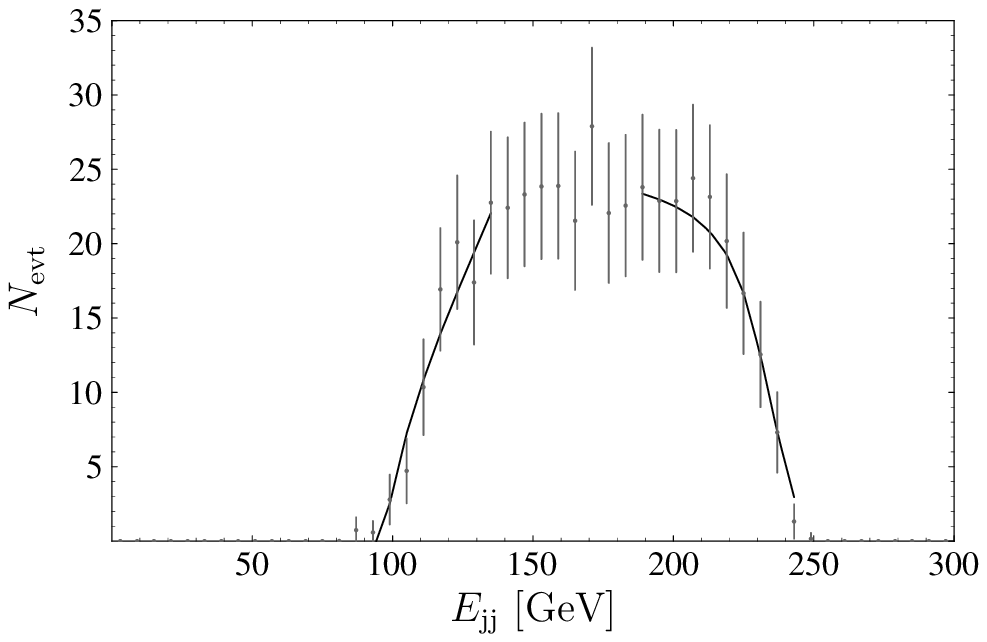, width=8.1cm} &
\epsfig{figure=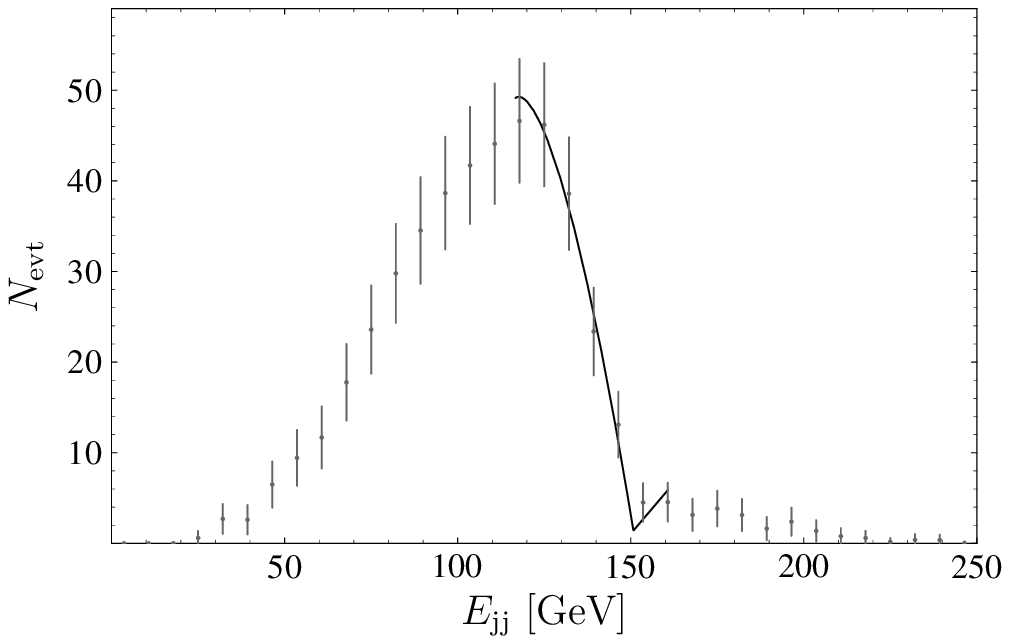, width=7.8cm} \\ 
(c) \\
\anc\hspace{1mm}%
\raisebox{-3.5cm}{\epsfig{figure=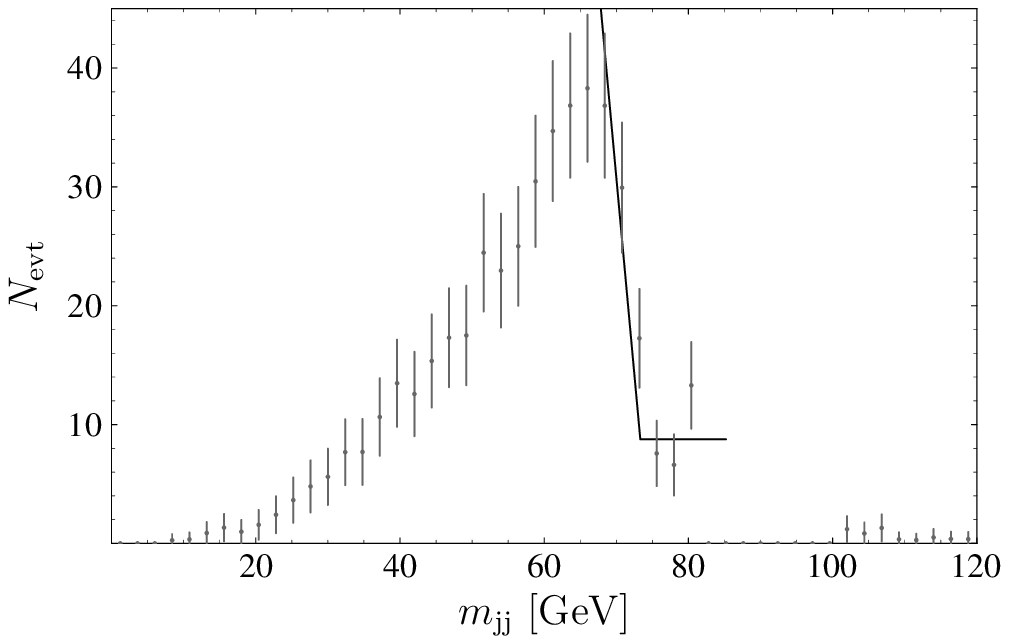, width=7.9cm}} &
\mycaption{(a) Fit to the energy spectrum of the jet pair originating from
$\neu_4$ decay, with $m_{jj} \sim \MZ$. (b) Fit to the energy spectrum of the
jet pair originating from
$\neu_2$ decay, with $m_{jj} \ll \MZ$. (c) With to invariant mass spectrum of
the jet pair from $\neu_2$ decay.}
\label{fg:neu24fit}
\end{tabular}
\end{figure}

As before, more information can be extracted from the decay distributions. The
energy spectra of the jet pairs stemming from the $\neu_2$ and $\neu_4$ decay
have distinct %upper and lower 
endpoints given by
\begin{equation}
\begin{aligned}
E_{\rm max,2} &= \frac{\mneu{2}^2 - \mneu{4}^2 - 2 \mneu{1}^2 \sqrt{s} +
s}{2\sqrt{s}}, 
\\
E_{\rm min,max,4} &= \frac{1}{4\mneu{4}^2\sqrt{s}} 
\Bigl ( \mneu{4}^4 - \mneu{2}^2\mneu{4}^2 + \mneu{4}^2\MZ^2 -
 \mneu{2}^2\MZ^2 + \mneu{4}^2 s + \MZ^2 s \\ & \qquad\qquad\qquad - 
 \mneu{1}^2 (\mneu{4}^2 -
 \mneu{2}^2 + s) \mp \sqrt{\lambda(\mneu{4}^2, \mneu{1}^2, \MZ^2) \,
   \lambda(\mneu{2}^2, \mneu{4}^2, s)}\Bigr ).
\end{aligned}
\end{equation}
In addition, the invariant mass of the jet pair from $\neu_2$ has a maximum
value corresponding to the mass difference of the neutralinos $\neu_2$ and
$\neu_1$,
\begin{equation}
m_{\rm jj,max,2} = \mneu{2} - \mneu{1}, \\
\end{equation}
All these kinematic edges can be fitted with simple functions, as shown in
Fig.~\ref{fg:neu24fit},
%%%%%%%%%%%%%%%%%%%%%%%%%%%%%%%%%%%%%%%%%%%%%%%%%%%%%%%%%%%%%%
%%%%%%%%%%%%%%%%%%%%%%%%%%%%%%%%%%%%%%%%%%%%%%%%%%%%%%%%%%%%%%
with the result
\begin{align}
\mneu{2} &= 106.6^{+12}_{-17} \gev, &  \mneu{4} &= 278.0^{+25}_{-18} \gev, &
\mneu{1} &= 33.3^{+12}_{-17}  \gev.
\end{align}
As for the case of $\neu_3\neu_4$ production, the errors on the neutralino
masses are relatively large.

\subsubsection{Chargino \boldmath $\cha^\pm_2$}

Heavy charginos $\cha_2^\pm$ can be produced in association with light
charginos $\cha^\pm_1$, $e^+e^- \to \cha^\pm_1\cha^\mp_2$.
While the $\cha^\pm_1$ mainly decays into $W$ bosons and the lightest
neutralino $\neu_1$, the $\cha_2^\pm$ has a large branching ratio of 62\% into
$Z W^\pm \neu_1$ via the decay chains $\cha^\pm_2 \to Z\cha^\pm_1 \to
Z W^\pm \neu_1$ and $\cha^\pm_2 \to W^\pm\neu_3 \to Z W^\pm \neu_1$.
These decay channels for the charginos lead to three gauge bosons and missing
energy  from the neutralinos. The gauge bosons themselves can decay through
various channels, but a good compromise between large statistics and clean
leptonic final states is obtained by requiring one $W$ boson to decay
leptonically, and the other $W$ and the $Z$ going into a hadronic final state.
The final state signature is then characterized by four jets, one lepton and
missing energy, $e^+e^- \to \cha^\pm_1\cha^\mp_2 \to Z \, W^+W^- \,
\neu_1 \, \neu_1  \to 4j\,l^\pm + \Eslash$. 
The signal can be enhanced by polarizing the beams with $P(e^+)$/$P(e^-)$ =
right/left, see Tab.~\ref{tab:polXsec}.

The most relevant SM backgrounds are triple gauge boson processes,
$e^+e^-\to W^+W^-Z$ and $t\bar{t}$ events. Production of heavy neutralino pairs
can also lead to three gauge bosons in the final state, and thus is another
important background.

After the standard cuts $p_t > 12$ GeV and
$\Eslash > 100$ GeV and a bottom veto, a number of additional selection cuts has to
be applied to reduce the backgrounds. Since the SM background, without neutralinos in the
final state, tends to give more energy to the gauge bosons, it can be reduced by
requiring the total hadronic energy to be $E_{\rm had} < 300$ GeV. The
invisible neutralinos in the signal also lead to a large value for the invariant
mass of the leptonic momentum and reconstructed missing 4-momentum,
$m_{l\scriptsize\Eslash}$, so that the cut $m_{l\scriptsize\Eslash} > 150$ GeV
reduces the SM further. For the SM background, this invariant mass can also be
reconstructed from the missing 3-momentum $\ppslash$ only, by assuming that
the missing energy originates from a neutrino, and it should be close the $W$
mass. Therefore the cut $|m_{l\scriptsize\pslash} - \MW| > 10$ GeV is also
effective on the background.
Furthermore, the signal is characterized by a large acoplanarity between the
lepton and combined jet system, so that the cut $\cos\phi_{\rm aco,lj} >
-0.7$ is useful.
Finally, two of the jets have to combine to the invariant mass of the $Z$ boson,
while the other two jets have to combine to $W$ mass,
$|m_{j_1j_2} - \MZ| < 10$ GeV and $|m_{j_3j_4} - \MW| < 10$ GeV. This
removes most of $\neu_2 \neu_4$ background and
is also effective on $t\bar{t}$.

After application of these cuts, the SM background is removed to a negligible
level, while still a sizeable contamination of background from $\neu_3\neu_4$
is left. In total $B=245$ background events remain, compared to $S= 186$ events
for the signal.
Since the cross-section for the neutralino process can be measured
independently, as described above, it can be subtracted, but the additional
error from this procedure needs to be taken into account. The resulting
expected precision for the $\cha^\pm_1\cha^\mp_2$ cross-section is
$\delta\sigma^\pm_{12} = 13$\%.

For the chargino signal, the spectrum of the 4-jet invariant mass has an upper
limit of $m_{\rm
inv,j,max} = \mcha{2} - \mneu{1}$, which can be used to extract information
about the heavy chargino mass. The neutralino background typically leads to
slightly smaller 4-jet invariant masses, so that this upper edge is not
contaminated.
From a fit to the data, one obtains
\begin{equation}
m_{\rm inv,j,max} = 287.2^{+5.4}_{-4.2} \gev,
\end{equation}
which together with the $\mneu{1}$ mass measurement from the
analysis of $\cha^+_1\cha^-_1$ production directly translates into
\begin{align}
\mcha{2} &= 319.5^{+5.5}_{-4.3} \gev. \label{eq:lcmc2}
\end{align}

\subsubsection{Combination of sparticle measurements at ILC}

Feeding in the precise measurement of the neutralino mass from the analysis
of $\cha^+_1\cha^-_1$ production, the masses of the heavier neutralinos from
$\neu_2\neu_4$ and $\neu_3\neu_4$ production can be determined much more
accurately,
\begin{align}
\mneu{2} &= 106.6^{+1.1}_{-1.3} \gev, &
\mneu{3} &= 181.5 \pm 4.9 \gev, 
&  \mneu{4} &= 278.0^{+2.5}_{-3.5} \gev.\label{eq:lcmn}
\end{align}
For the lightest neutralino and the charginos, the expected errors given in the previous 
sections are not improved by combining with the other neutralino observables, 
so that one obtains
\begin{align}
\mneu{1} &= 33.3^{+0.4}_{-0.3} \gev, &
\mcha{1} &= 164.98 \pm 0.05 \gev, 
&  \mneu{4} &= 319.5^{+5.5}_{-4.3} \gev.\label{eq:lcmc}
\end{align}
From a $\chi^2$  fit to all mass and cross-section observables,
constraints on the underlying neutralino and chargino parameters can be
extracted. For completeness, we also allow a cubic singlet coupling $\kappa$ as
in the NMSSM. In the nMSSM, $\kappa$ must be zero, but it is interesting not to
impose this requirement a priori, but see how well it can be checked from an
experimental analysis. The parameter $\kappa$ enters in the (5,5)-entry of the
neutralino mass matrix,
\begin{equation}
M_{\neu} = \begin{pmatrix}
M_1 & 0 & -c_\beta\sw\MZ & s_\beta\sw\MZ & 0 \\
0 & M_2 & c_\beta\cw\MZ & -s_\beta\cw\MZ & 0\\
-c_\beta\sw\MZ & c_\beta\cw\MZ & 0 & \lambda \vs & \lambda v_2 \\
s_\beta\sw\MZ & -s_\beta\cw\MZ & \lambda \vs & 0 & \lambda v_1 \\
0 & 0 & \lambda v_2 & \lambda v_1 & \kappa \\
\end{pmatrix},
\end{equation}
The possible measurements at the ILC analyzed here comprise mass measurements
for four neutralino and two chargino states, as well as four cross-section
measurements. They can be used to derive bounds on the seven unknown parameters
in the neutralino and chargino mass matrices. Furthermore, the cross-section
measurements also allow to place limits on the masses of the sneutrino and
selectron, which appear in the t-channel of the chargino and neutralino
production diagrams.
Based on the analysis of the expected experimental error in the previous
subsections, the following constraints on the underlying parameters are
obtained:
\begin{align}
M_1 &= (122.5 \pm 1.3) \gev, & |\kappa| &< 2.0 \gev, & 
 m_{\tilde{\nu}_{\rm e}} &> 5 \tev, \nonumber
\\
M_2 &= (245.0 \pm 0.7) \gev, & \tan\beta &= 1.7 \pm 0.09, &
 \mseR &> 1 \tev. \label{eq:ilcres}
\\
|\lambda| &= 0.619 \pm 0.007, & |\phi_{\rm M}| &< 0.32, \nonumber
\\
v_{\rm s} &= (-384 \pm 4.8) \gev, \nonumber
\end{align}
The extraction of the parameters $\lambda$ and $v_{\rm s}$ is strongly
correlated, which can be understood by the fact that these parameters enter in
the chargino and neutralino mass matrices mainly through the combination $\mu =
-\lambda v_{\rm s}$. As a consequence, the effective parameter $\mu$ itself is
determined more precisely than $v_{\rm s}$, $\mu = (238 \pm 1.2) \gev$.

The results of the fit show that the sizable value of the trilinear
Higgs coupling $\lambda$ can be established, which is a necessary
requirement to avoid the Higgs mass bounds and allow a
successful baryogenesis in singlet extensions of the MSSM.
Furthermore, a strong upper bound on the value of $\kappa$ is obtained, which
allows a distinction between the two typical types of singlet extensions, the
NMSSM and the nMSSM.

%%%%%%%%%%%%%%%%%%%%%%%%%%%%%%%%%%%%%%%%%%%%%%%%%%%%%%%%%%%%%%

\section{Cosmological implications}
\label{sc:cosmo}

%%%%%%%%%%%%%%%%%%%%%%%%%%%%%%%%%%%%%%%%%%%%%%%%%%%%%%%%%%%%%%

The cosmological energy density of the main components of matter, baryons 
and dark matter, is measured with a remarkable precision 
\cite{Spergel:2006hy}.  In units of the critical density%
\footnote{$\rho_c = 3 H_0^2/(8 \pi G_N)$ where $H_0 = h \times 100~km/s/Mpc$ 
is the present value of the Hubble constant, $h = 0.709^{+0.024}_{-0.032}$, 
and $G_N$ is Newton's constant.}
\begin{eqnarray}
& \Omega_{\rm B} h^2   = 0.02233^{+0.00124}_{-0.00172}, & \nonumber \\
& \Omega_{\rm CDM}h^2 = 0.1106^{+0.0113}_{-0.0151}, & \nonumber 
\end{eqnarray}
at 95\%~CL.
According to the observations, the baryon density is dominated by 
baryons while anti-baryons are only secondary products from high energy 
processes.  The source of this baryon--anti-baryon asymmetry and the nature 
of dark matter are major puzzles of particle and astrophysics.

Assuming that inflation washes out any initial baryon asymmetry after the 
Big Bang, a dynamic mechanism should generate the asymmetry after inflation. 
Most microscopic mechanisms for baryogenesis fulfill the three Sakharov 
requirements:
\begin{itemize}
  \item baryon~number~($B$)~violation,
  \item CP~violation, and
  \item departure~from~equilibrium.
\end{itemize}
These conditions are satisfied in the supersymmetric extensions of the 
Standard Model during the electroweak phase transition. This is the 
basis for electroweak baryogenesis (EWBG). 
Baryon number violation occurs in the nMSSM due to quantum transitions 
between inequivalent SU(2) vacua that violate the sum of baryon and lepton 
number $B + L$.  When the electroweak phase transition is first order, 
bubbles of broken phase nucleate within the symmetric phase as the Universe 
cools below the critical temperature.  These provide the necessary departure 
from equilibrium.  To generate the observed baryon asymmetry the electroweak 
phase transition has to be strongly first order.  For light Higgs bosons, a 
first order phase transition can be induced by the trilinear coupling 
between the singlet and Higgs fields \cite{gaga}.  Such a term produces a 
sufficient contribution to the cubic term in the one-loop effective 
potential that is responsible for making the phase transition first order.

As discussed in Section \ref{sc:nmssm} the lightest superpartner in the 
nMSSM is typically the lightest neutralino, with a mass below about 60 GeV (for
$\lambda$ below the perturbative bound) and a sizeable singlino 
component.  Cosmologically this signals danger since a light, stable 
particle with very weak gauge couplings may have a relic abundance 
substantially above the observed amount.  But if the lightest neutralino is 
able to annihilate sufficiently well, this particle is a good dark matter 
candidate \cite{Olive:1990aj}.

For values of $\tan\beta$  and $\lambda$ consistent with the perturbative 
limit, the lightest neutralino also acquires some higgsino component.  Since 
its light mass allows for efficient annihilation nearby the $Z^0$ resonance, 
its relic density spans a wide range as the function of its mass.  While the 
s-channel $Z^0$ exchange is the dominant annihilation mode, there are also 
contributions from s-channel Higgs boson exchanges generated by the 
trilinear singlet-Higgs-Higgs term of the superpotential.  These 
contributions tend to be significant only near the corresponding mass poles 
\cite{gaga}.  

When the neutralino relic density is consistent with the astrophysical 
constraints, and the model parameters with electroweak baryogenesis and the 
perturbative bound, the next lightest supersymmetric particle (NLSP) is always at 
least 15 percent (and frequently more than 25 percent) heavier than the LSP, 
assuring that the co-annihilation contribution is strongly 
Boltzmann-suppressed.  
Moreover, since the NLSP is typically a neutralino with large bino component,
which has small couplings to the gauge bosons,
the co-annihilation cross section between the 
LSP and NLSP is suppressed even further and does not play any role in the following 
discussion.

\subsection{Dark matter relic density}

The relic abundance of neutralinos is computed numerically by solving the 
Boltzmann equation for the number density of the supersymmetric particles. 
The complex phase of $\lambda$ enters our relic density calculation directly 
via the couplings and indirectly through the masses of the neutralinos and 
charginos.  After diagonalizing the gaugino, sfermion and Higgs mass 
matrices, we calculate the annihilation cross sections with complex 
couplings.  In doing this, we follow techniques used in 
Refs.\cite{Gondolo:2004sc,Katsanevas:1997fb}.  The co-annihilation processes 
are checked to contribute insignificantly to the final result.

%%%%%%%%%%%%%%%%%%%%%%%%%%%%%%%%%%%%%%%%%%%%%%%%%%%%%%%%%%%%%%
\begin{figure}[tb]
\centering
\epsfig{figure=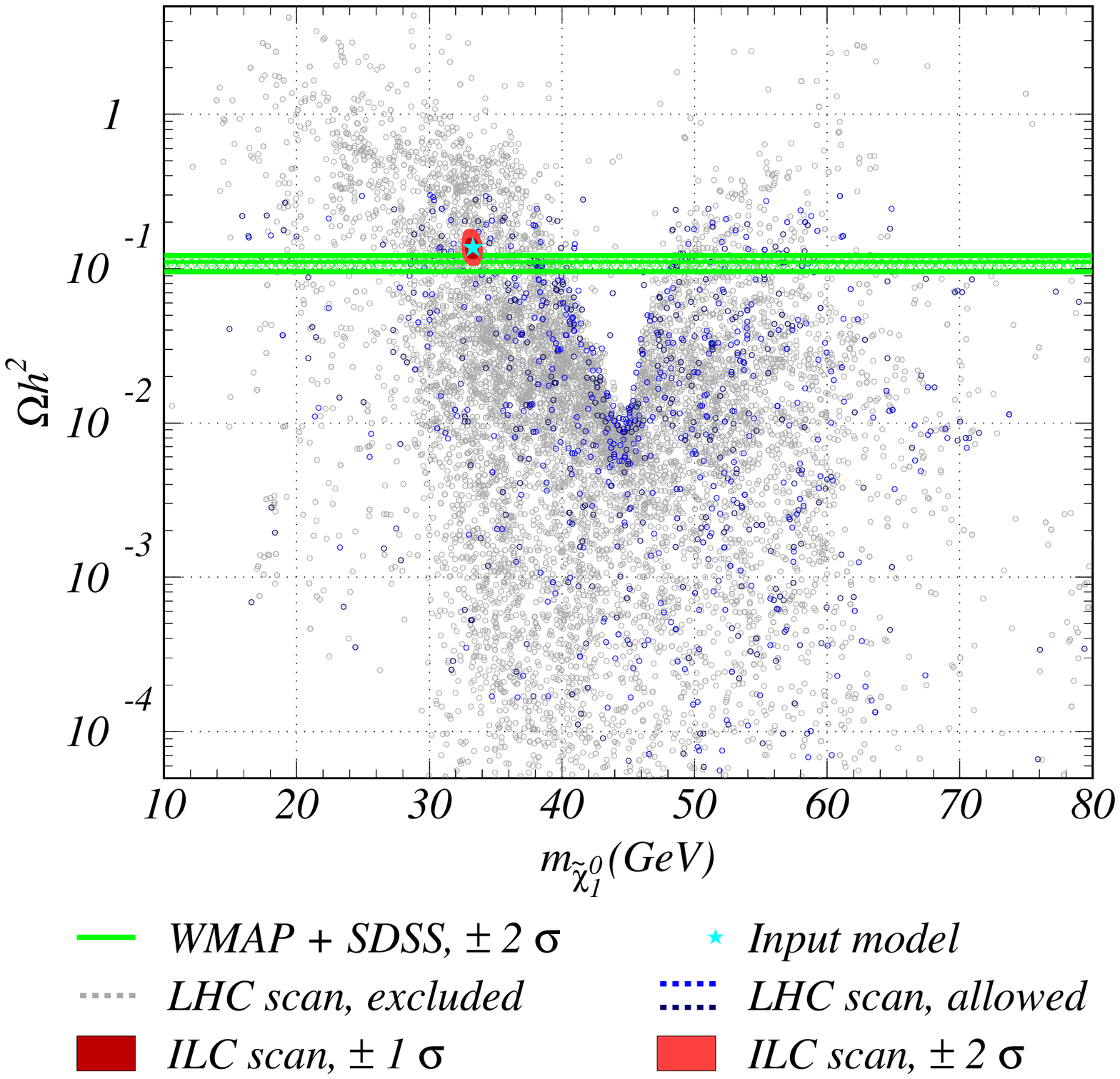, width=0.49\textwidth}
\epsfig{figure=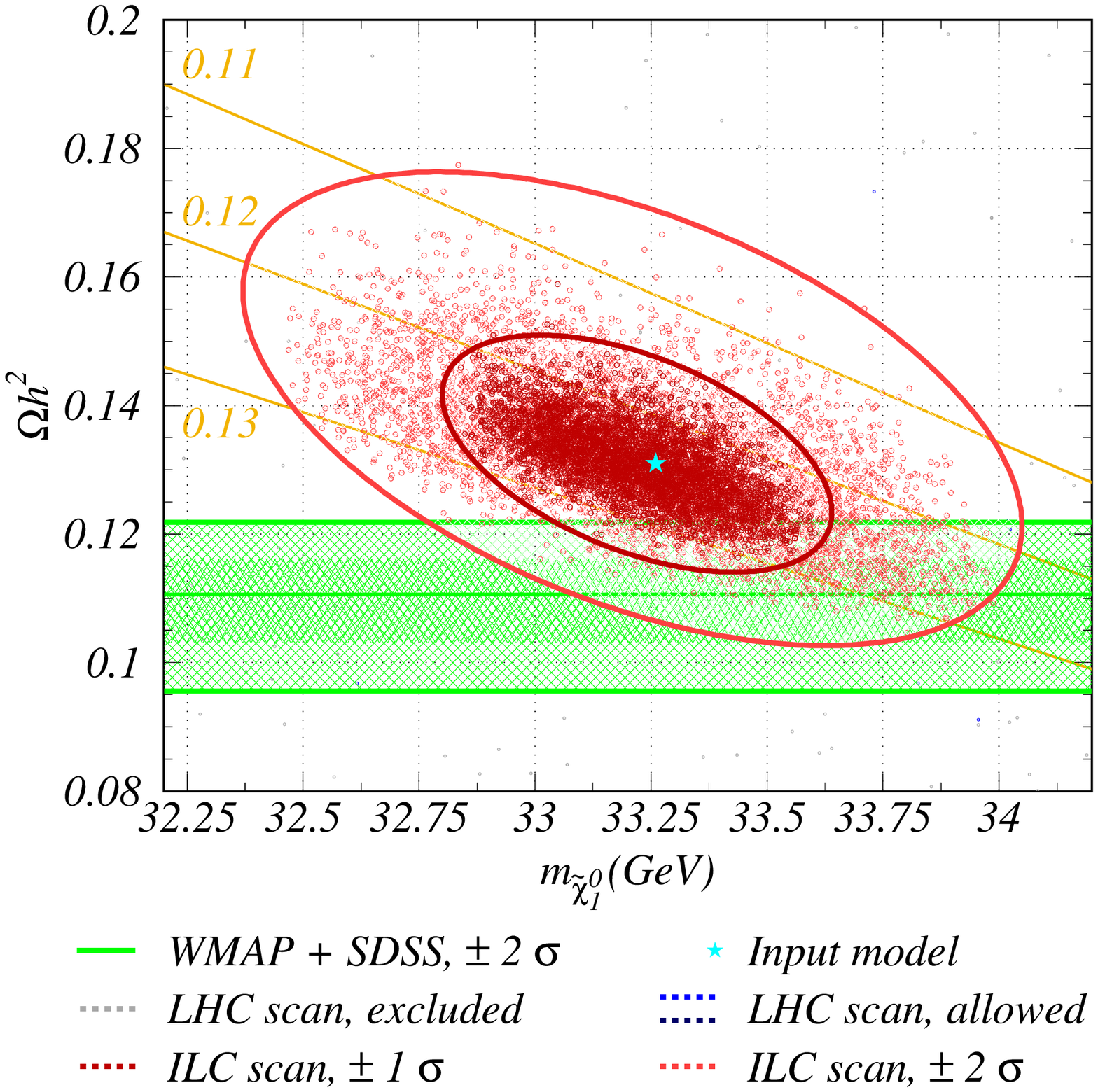, width=0.49\textwidth}
\mycaption{Neutralino relic density as the function of the neutralino mass.
 Dark (light) blue dots represent the 1 (2) $\sigma$ precision of
 the LHC determination of $\Omega h^2$, while gray dots 
would be allowed by LHC data, but are excluded by current low-energy and
astrophysical bounds.
Red dots show the expected ILC precision for 
the examined model point.  The present WMAP and SDSS combined 2 $\sigma$ 
limits are shown by the green shaded band.  The right frame shows the ILC
scan in more details, with contours of constant values of the mixing parameter
$\left(|N_{14}|^2 - |N_{13}|^2\right)$ indicated by the yellow lines.}
\label{Fig:oh2_mz1_lhc}
\end{figure}
%%%%%%%%%%%%%%%%%%%%%%%%%%%%%%%%%%%%%%%%%%%%%%%%%%%%%%%%%%%%%%

After superpartners are discovered and their properties being measured at 
colliders one has to assure the consistency of the collider and 
astrophysical data.  A crucial part of this is to ensure that the lightest, 
stable supersymmetric particle provides a reasonable amount of the observed 
cold dark matter.  

As discussed before, the LHC will restrict some of the soft supersymmetric 
parameters within certain ranges.  Using these ranges, we can calculate the 
possible amount of neutralino dark matter, $\Omega h^2$, within the given 
supersymmetric model.  In this section we use our results obtained above for 
scenario A.  To obtain an estimate of the precision the LHC can determine 
$\Omega h^2$ in the nMSSM, we randomly sampled the nMSSM parameter space in 
the following parameter region:
\begin{eqnarray}
&& 0 < M_1 < 200 {\rm ~GeV}, ~ 100 {\rm ~GeV} < M_2 < 300 {\rm ~GeV}, ~
   0 < |\lambda| < 1, ~ -\pi < \phi_{\rm M} < \pi, \nonumber \\
&& -1000  {\rm ~GeV} < \vs < -100 {\rm ~GeV}, ~ -1000  {\rm ~GeV} < \kappa < 1000 {\rm ~GeV}, ~
   0 < \tan\beta < 30.  \quad
\end{eqnarray}
Additionally, for the first generation sleptons, we use the following ranges
both in our LHC and ILC scans:
\begin{equation}
0.5 {\rm ~TeV} < M_{e_R} < 10 {\rm ~TeV}, \quad 0.5  {\rm ~TeV} < M_{L_1} < 10 {\rm ~TeV}.
\end{equation}
The results of this scan, within the parameter region that would be allowed by
LHC measurements at the 1~$\sigma$ level, are projected on the lightest
neutralino relic density vs. mass plane  in Figure \ref{Fig:oh2_mz1_lhc}.

Blue (gray) dots represent models that are (not) allowed by either the WMAP,
the electron EDM, or WIMP direct detection limits.  The dark (light) blue
dots show the 1 (2) $\sigma$ precision of the LHC determination of $\Omega h^2$.
All blue points satisfy all known low energy, collider and
astrophysical constraints.  Among these, the most stringent bounds come from
the LEP and Tevatron Higgs and sparticle mass limits, WMAP, and the
direct WIMP detection experiments.

Fig.~\ref{Fig:oh2_mz1_lhc} (left) clearly demonstrates that from LHC
measurements no
meaningful constraints on $\Omega h^2$ can be derived.  In contrast, as
depicted in Fig.~\ref{Fig:oh2_mz1_lhc} (right), the ILC 
1 $\sigma$ precision, indicated by the dark red dots, is comparable to the
present WMAP and SDSS combined 2 $\sigma$ limits, shown by the green shaded
band. Even at the 2 $\sigma$ level, the ILC could provide important information.
In more quantitative terms the constraints on the neutralino relic density from 
the ILC measurements are 
\begin{equation}
\boldmath 0.105 < \Omega h^2 < 0.178.	\qquad (2 \sigma)
\end{equation}
As can be seen from Fig.~\ref{Fig:oh2_mz1_lhc}, the computed dark matter
density is strongly correlated with the lightest neutralino mass. Therefore a
precise measurement of $\mneu{1}$ is essential for the accurate prediction of
$\Omega_{\rm CDM}$. As shown in the previous chapter, for the sample scenario A
the neutralino mass can be constrained best from chargino observables, in
particular using a chargino threshold scan.

Once the neutralino mass from the ILC and their relic abundance from 
astrophysical measurements are known with the precision indicated by 
Fig.~\ref{Fig:oh2_mz1_lhc}, we can even extract information on the 
neutralino mixing matrix.  This is possible because the 
relic density of neutralinos depends only on the neutralino mass and 
the $Z\neu_1\neu_1$ coupling, given by eq.~\eqref{GZn1n1}. 
Constant contours of $\left(|N_{14}|^2 - 
|N_{13}|^2\right)$ in Fig.~\ref{Fig:oh2_mz1_lhc} indicate the precision at 
which this coupling can be extracted from the combination of collider and 
astrophysical data in the ILC era.
While in principle it is possible to determine the neutralino mixing matrix completely
from ILC data alone by reconstructing the neutralino mass matrix, the combination
of collider and astrophysical data allows to extract information about the mixings without
having to assume the particle content of the nMSSM/NMSSM. This can provide an 
interesting cross-check about the structure of the underlying model.

The accuracy in the interplay of the collider and astrophysical measurements
can be considered spectacular. If a consistency of the measurements as shown in
Fig.~\ref{Fig:oh2_mz1_lhc} (right) were demonstrated it would be an impressive
test of this model. On the other hand, it is worth noting that an inconsistency
of the ILC and astrophysical data would pose an interesting, and potentially 
fruitful, situation.  A collider indication of higher than WMAP or PLANCK 
allowed dark matter density, for example, may signal non-standard cosmology. 
Such an inconsistency could be the first sign that the temperature of the 
radiation dominated epoch before Big-Bang nucleosynthesis is not high enough for neutralinos to 
reach kinetic and chemical equilibrium.  Alternatively, the entropy of 
matter and radiation may not be conserved or late production of entropy tips 
its balance.  
If instead the relic density implied by collider data were too small then this
could be interpreted as a hint of additional dark matter constituents besides
the neutralino. In either case, the ILC has a potential to uncover far 
reaching cosmic connections.

%%%%%%%%%%%%%%%%%%%%%%%%%%%%%%%%%%%%%%%%%%%%%%%%%%%%%%%%%%%%%%

\subsection{Dark matter direct detection}
\label{sc:direct}

%%%%%%%%%%%%%%%%%%%%%%%%%%%%%%%%%%%%%%%%%%%%%%%%%%%%%%%%%%%%%%
\begin{figure}[tb]
\centering
\epsfig{figure=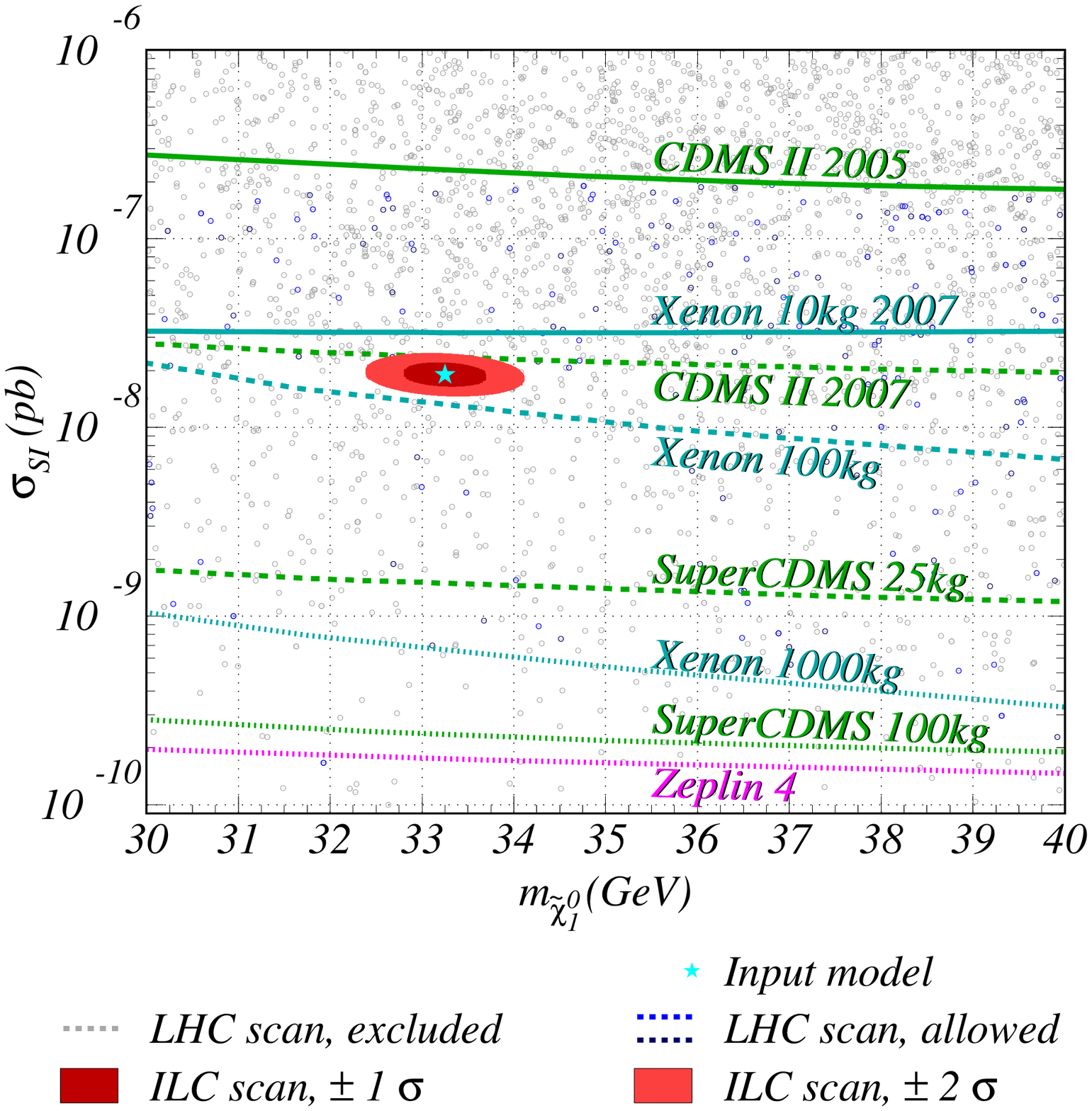, width=0.48\textwidth}
\epsfig{figure=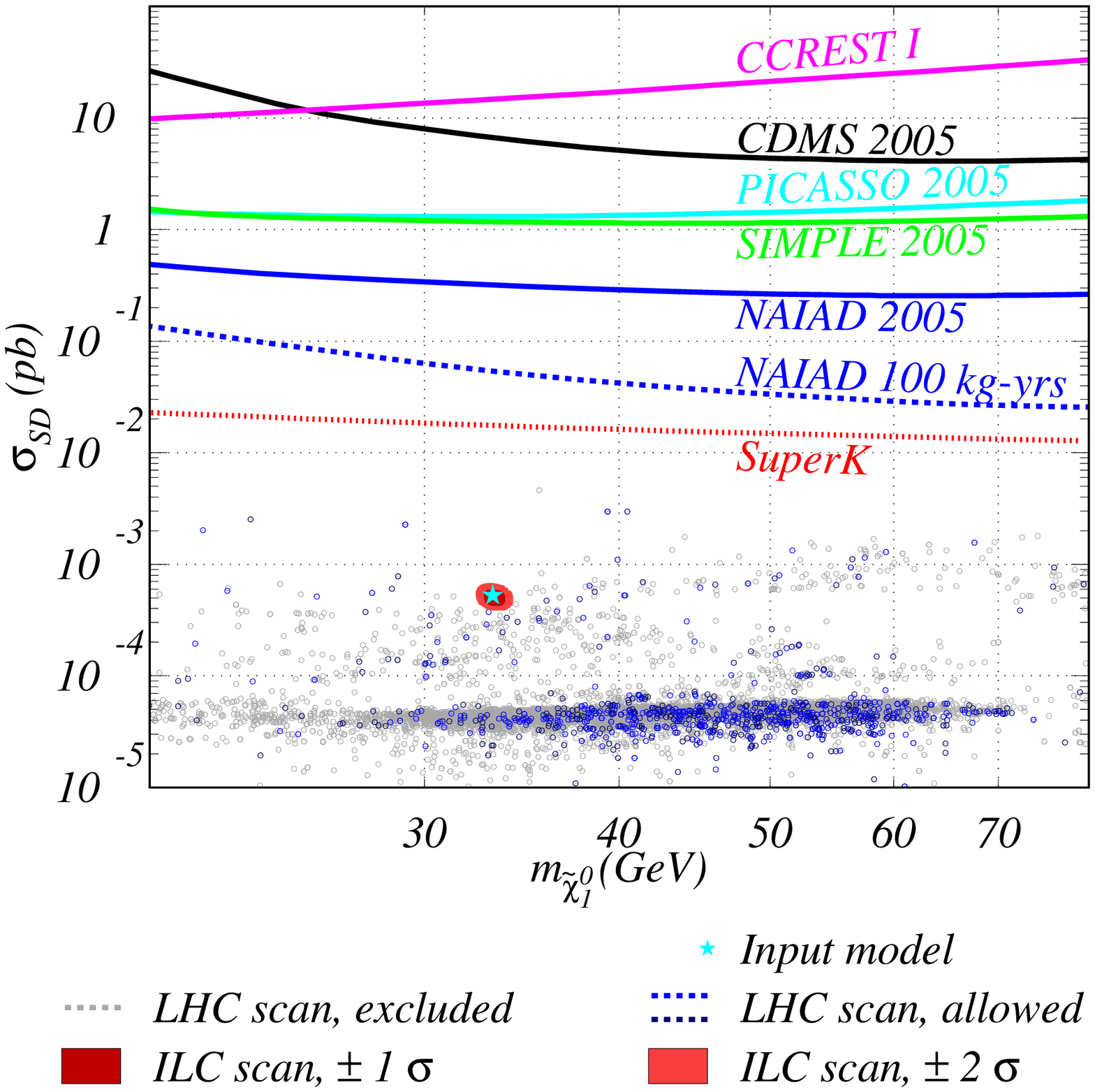, width=0.48\textwidth}
\mycaption{Left frame: Spin independent neutralino-proton elastic scattering 
cross sections as a function of the (lightest) neutralino mass.  Blue and 
gray dots show the results of the LHC scan projected to this plane.  Red 
dots correspond to the ILC scan.  The regions above the various curves are 
excluded by the indicated experiments. Right frame: Spin dependent 
neutralino-proton elastic scattering cross section compared to various 
direct detection experiments, and to SuperK indirect limits for neutrinos 
stemming from neutralino annihilation in the Sun.  The red region indicates 
the prediction from ILC measurements.}
\label{Fig:ssi_mz1}
\end{figure}
%%%%%%%%%%%%%%%%%%%%%%%%%%%%%%%%%%%%%%%%%%%%%%%%%%%%%%%%%%%%%% 

Another important requirement is the consistency of collider and dark matter 
direct detection data.  Direct detection experiments search for weakly 
interacting massive particles via their elastic scattering off nuclei by 
measuring the nuclear recoil.  There are numerous existing and future 
experiments engaged in this search \cite{Gaitskell:2004gd}.  These 
experiments uniformly express their observations in terms of the 
neutralino-proton scattering cross section.  Most of the current direct 
detection experiments are primarily sensitive to scalar interactions of 
neutralinos with nuclei, and typically the most stringent limits are the 
spin independent ones.  This is because for heavy nuclei the spin 
independent cross section, being proportional to the squared mass of the 
target nucleus, is highly enhanced compared to the spin dependent one.

While both spin dependent and independent scattering rates receive 
contributions from s-channel squark exchanges, the dominant processes are 
t-channel $Z$ boson and Higgs exchange %(neglecting loop diagrams) 
for spin dependent and independent interactions, respectively.  Due to close to 
resonant $Z$-exchange the spin dependent scattering cross section is 
enhanced in the nMSSM. However, as a result of the small $Z\neu_1\neu_1$ coupling,
the cross section will not be within reach of
the next generation of direct detection experiments.

Fig.~\ref{Fig:ssi_mz1} shows the neutralino-proton elastic scattering cross
sections in the nMSSM as the function of the  (lightest) neutralino mass.  The
results of our LHC (blue and gray dots) and  ILC (red dots) scans are projected
to this plane, and the regions above the  various curves are excluded by each
indicated experiment.  Due to the large  singlino component in $\neu_1$, the
spin dependent  neutralino-proton elastic scattering cross section is very
small. The spin independent cross section, on the other hand, is enhanced
through the sizeable scalar self-coupling $\lambda$, which generates the
dominant contribution for the coupling of $\neu_1$ to the CP-even Higgs bosons.
As the left frame of Fig. 9 shows, the current and next generation direct
detection experiments can already test what role the nMSSM plays in the
explanation of dark matter, see also Ref.~\cite{Barger:2007nv}. In particular, 
SuperCDMS and Xenon should be able to discover or rule out benchmark point A in
the near future.

The right frame of Fig. 9 shows the results of the scan for the spin dependent
neutralino-proton elastic scattering cross section $\sigma_{SD}$.  The
sensitivity of many spin dependent direct detection experiments
\cite{Gaitskell:plotter}, together with the SuperK limit is also shown.
Contrary to the spin independent case, the present and near future direct
detection experiments fall short of observing this scenario.  The strongest
limit on $\sigma_{SD}$ presently comes from the SuperK search for an excess of
high energy muon neutrinos from WIMP annihilation in the Sun.  Since, among
other astrophysical quantities, this neutrino flux depends on $\sigma_{SD}$,
the SuperK measurement is able to constrain it from above \cite{Desai:2004pq}.
In the future, the most stringent bounds on $\sigma_{SD}$ may also come from
indirect detection experiments. Unfortunately, the sensitivity of the IceCUBE
neutrino telescope has only sensitivity for energies above the maximum energy
of neutrinos expected from the nMSSM scenario, so we do not expect a strong
constraint from it.

One should keep in mind that in general the direct detection cross sections
depend on the CP-violating gaugino phase $\phi_{\rm M}$, but at present there
are large theoretical uncertainties associated with the determination of the
value of $\phi_{\rm M}$ preferred for baryogenesis. The choice $\phi_{\rm M} =
0.14$ in the benchmark scenario A serves as a guideline, but the correct value
might differ by a factor of two. When changing the phase to the rather extreme
case $\phi_{\rm M} \approx \pi$, the spin
independent cross section increases moderately by roughly 50\%. 
Thus our main conclusions, that $\sigma_{SI}$ can be tested in the
near future, while  $\sigma_{SD}$ is likely out of reach of all planned
experiments, remain the same irrespective of the value of $\phi_{\rm M}$.

The relatively small dependence of the direct detection cross section on the
phase $\phi_{\rm M}$ of the gaugino mass parameters can be undertstood by
observing that the lightest neutralino is mainly singlino and has only a small
gaugino component. As pointed out in section~\ref{sc:cp}, however, a relevant
phase can also appear in the singlet--Higgs sector, for example in the
parameter $a_\lambda$. Such a complex phase could have a much stronger impact
on the direct detection scattering cross sections, depending on the value of
$\arg(a_\lambda)$. It would be interesting to study this possibility and its
interrelation with the generation of baryon asymmetry. We reserve such a study
for a future investigation.

%%%%%%%%%%%%%%%%%%%%%%%%%%%%%%%%%%%%%%%%%%%%%%%%%%%%%%%%%%%%%%

\subsection{Baryogenesis}

As shown in the previous section, the measurements of the chargino 
and neutralino sector at
the LHC and the ILC  provide a test of the presence of
light charginos and neutralinos, necessary to generate 
the dark matter relic density. In order to probe the mechanism of electroweak
baryogenesis with collider results, two conditions need to be tested:
the type of the electroweak phase transition must be strongly first order, and
there must be CP violating processes active during this phase transition.

In our benchmark scenario, CP violation is introduced in the baryon-number
generating processes through light chargino currents. For this mechanism, the
charginos need to be light enough so that they are not decoupled at the
temperature of electroweak symmetry breaking. Using the experimental
results from LHC and ILC, see eqs.~\eqref{eq:lhcres} and \eqref{eq:ilcres},
the existence of sufficiently light chargino can easily be tested.
In addition, the presence of a complex CP-violating parameter in the chargino
sector is required for baryogenesis. However, even with the high precision of
ILC, only an upper bound on the phase $\phi_{\rm M}$ of the gaugino mass
parameters can be obtained, see eq.~\eqref{eq:ilcres}.

To test the other condition, the strength of the first order phase transition,
the Higgs sector of the model needs to be analyzed. The strength of the phase
transition
can be calculated from the effective Higgs potential, see e.g.~\cite{gaga}.
It depends crucially on the supersymmetry breaking term $\ms^2$ and
$a_\lambda$, which are not constrained by the analysis of charginos and
neutralinos. However, as we will show below, information
about these parameters may be obtained by the precise determination of
the CP-even Higgs boson masses, which would be possible at the ILC. 

From the condition of the strongly first order phase transition,
one finds the following conditions on the parameters of the Higgs
potential~\cite{gaga}: 
\begin{align}
\ms^2 &= -a_\lambda v_1 v_2 / \vs - \tad / \vs - \lambda^2 v^2 \nonumber \\
&\in \{ (50 \gev)^2, (200 \gev)^2 \} \qquad \text{for perturbative } \lambda
\lesim 0.8, \\
|D| &\equiv \left| \frac{1}{\ms^2\sqrt{\lambda^2/4 \, \sin^2 2\beta + 
\bar{g}^2/8 \, \cos^2
2\beta}} \left( \lambda^2 \tad/\ms - a_\lambda
\sin\beta \cos\beta \; \ms \right) \right| \gesim 1,
\end{align}
where we have introduced the quantity $D$ for abbreviation.

As stressed above,
constraints on these parameters can be obtained from the Higgs masses. 
To relate the masses to the underlying parameters, the Higgs mixing matrices
need to be reconstructed.
Following the discussion in section~\ref{sc:cp}, we have assumed
CP-conservation in the Higgs sector, to that mixing occurs only between Higgs
boson with the same CP quantum numbers.
The heavy Higgs states $S_3, P_2, H^\pm$ with masses of the order of $\MA$ are
not within reach of either the LHC or ILC (the most promising discovery
channels for heavy Higgs bosons at the LHC are suppressed due to the small
value of $\tan\beta$ in our scenario). From the fact that the charged Higgs
boson would not be observed at the ILC with $\sqrt{s} = 1 \tev$, one could
derive the limit $\MA > 500$ GeV. Due this bound the heavy CP-even Higgs
boson is essentially decoupled, so that the lightest CP-even Higgs has only
sizeable mixing with the
second-lightest CP-even Higgs. Both of these masses can be measured at ILC.
The mass matrix is
\begin{equation}
M_{\rm S_{1,2}}^2 =
\begin{pmatrix}
\MZ^2 \cos^2 2\beta + \lambda^2 v^2 \sin^2 2\beta &
v (a_\lambda \sin 2\beta + 2 \lambda^2 \vs) \\
v (a_\lambda \sin 2\beta + 2 \lambda^2 \vs) &
\ms^2 + \lambda^2 v^2 
\end{pmatrix} + \Delta M_{\rm S_{1,2}}^2,
\label{eq:ms}
\end{equation}
where $\Delta M_{\rm S_{1,2}}^2$ represents radiative corrections. The largest 
corrections stem from top-stop loops, but they are relatively small except for the (1,1)-entry,
\begin{align}
\Delta M_{\rm S_{1,2}}^2 \equiv
\begin{pmatrix}
\Delta_{S11} & \Delta_{S12} \\ \Delta_{S21} & \Delta_{S22} 
\end{pmatrix}
&\approx
\begin{pmatrix}
\Delta_{S11} & 0 \\ 0 & 0
\end{pmatrix},
&
\text{with\ \ }
\Delta_{S11} &\approx \frac{3}{8\pi^2} \frac{\mt^4}{v^2} \, \log 
\frac{m_{\tilde{t}_1}^2 m_{\tilde{t}_2}^2}{\mt^4}. \label{eq:dms}
\end{align}
A rough constraint on $\ms$ can be obtained by considering that one of the two
eigenvalues of the mass matrix is always larger than the diagonal entries of the
matrix, while the other eigenvalue is smaller than the diagonal entries. Ergo:
\begin{equation}
M_{S1}^2 < \ms^2 + \lambda^2 v^2 %+ \Delta_{S22} 
< M_{S2}^2.
\end{equation}
Using the uncertainties for $\lambda$ %and $\tan\beta$ 
from the fit to the
ILC measurements in the neutralino sector, one gets
\begin{equation} 
(41 \pm 3)^2 \gev^2 \lesim \ms^2 \lesim (114 \pm 1)^2 \gev^2,
\end{equation}
compared to the input value of $\ms = 106.5$ GeV. Thus this very crude estimate
is already sufficient to establish $\ms$ in the required range.

A more detailed determination would be possible by looking at the whole matrix
eq.~\eqref{eq:ms}, including full one-loop corrections \cite{hcorr}, not only
the
leading term in eq.~\eqref{eq:dms}. The radiative corrections add 
an additional uncertainty due to the parametric
dependence on the stop masses $m_{\tilde{t}_{1,2}}$, $A_{\rm t}$ and $M_{\rm
A}$. 

These parameters would need to be constrained from experiment.
While in our scenario the stops are too heavy to be produced at a 1 TeV linear
collider, one can try to search for them at the LHC. The study of
Ref.~\cite{nostop} finds that a signal from decays of gluinos into stops can be
identified with a dedicated analysis. Ref.~\cite{nostop} also proposes a
strategy to measure the stop mass, although a translation to our scenario is
not straightforward. Nevertheless, to exemplify the improvement that such a stop
analysis could bring for the understanding of the Higgs sector, we here
simply assume that the stop masses can be measured with an error of 
$\delta m_{\tilde{t}} = 50$ GeV. For the parameter $A_{\rm t}$  the situation is
more difficult, since it cannot be measured
directly. However, given that in our scenario there is only a relatively small
difference between the two stop masses, which we assume can be measured, one can
infer that $A_{\rm t} \lesim 500 \gev$. As far as $\MA$ is concerned, only a
lower limit of 500 GeV could be obtained in our scenario, as pointed out above.
With these constraints, and taking into account the
expected errors for all relevant masses and parameters, the full one-loop
analysis yields
\begin{align}
a_\lambda &= (373^{+17}_{-21}) \gev, &  \ms &= (106 \pm 18) \gev, \nonumber \\
 \tad^{1/3} &= (156^{+25}_{-39}) \gev, & |D| &\sim 1.0 \pm 0.65. 
 \label{eq:bgres}
\end{align}
Note that the parameters $a_\lambda$ and $\ms$ can be constrained very
precisely from the measurement of Higgs masses at the ILC. On the other hand,
the necessary condition $|D|>1$ cannot be proven with sufficient precision,
although the result in eq.~\ref{eq:bgres} is consistent with this condition.

In summary, measurements at future colliders can allow us to establish the
chargino and Higgs mass parameters to be in the range required for electroweak
baryogenesis in the nMSSM, but they do not seem sensitive enough to yield 
definitive answers to the questions of the first order phase transition and of
the presence of additional CP violation.

\subsubsection{Electron Electric Dipole Moment}

%%%%%%%%%%%%%%%%%%%%%%%%%%%%%%%%%%%%%%%%%%%%%%%%%%%%%%%%%%%%%%
\begin{figure}[tb]
\centering
\epsfig{figure=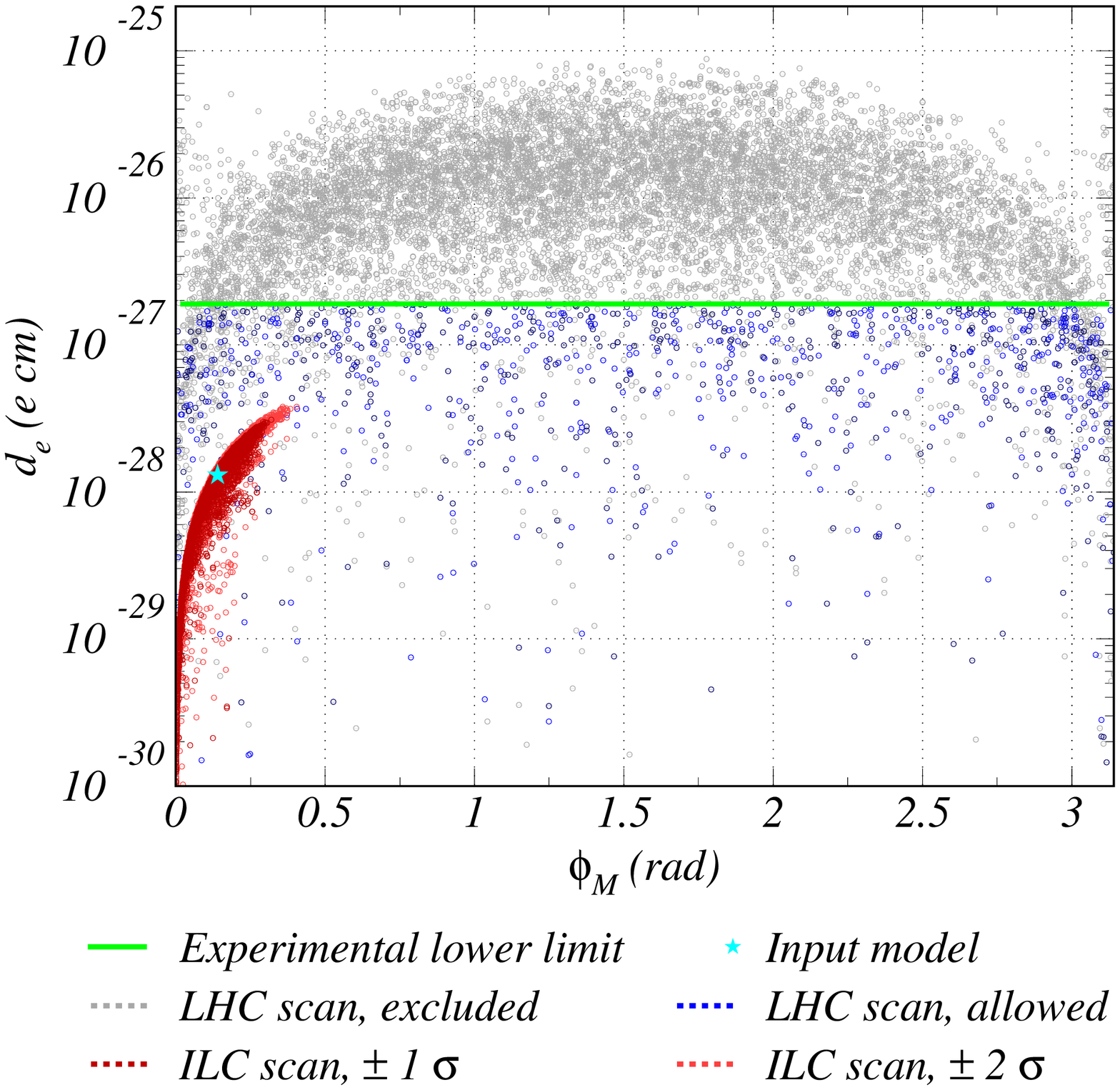, width=9cm}
\mycaption{Comparison of the current bound on the electron electric dipole
moment with parameter regions allowed by expected LHC and ILC measurements for
the scenario A. The results are given as a function of the complex phase 
$\phi_{\rm M}$.}
\label{fg:edm}
\end{figure}
%%%%%%%%%%%%%%%%%%%%%%%%%%%%%%%%%%%%%%%%%%%%%%%%%%%%%%%%%%%%%%

A necessary requirement of the electroweak baryogenesis scenario is the 
presence of non-vanishing CP-violating phases in the chargino--neutralino 
sector. In this work, we have assumed that these phases are associated with 
the gaugino sector of the theory. However, the collider, or the dark matter 
constraints described in the previous sections are not sufficient to 
determine the exact value of the CP-violating phases necessary for the 
generation of the baryon-antibaryon asymmetry. An important question is if 
one could obtain information about these phases from the measurement of, for 
instance, the electron electric dipole moment. It is advantageous to use the 
electron EDM since it is precisely measured, has relatively low theoretical 
uncertainties, and for the phases relevant to the model under study gives 
the strongest constraint. Since both the baryon asymmetry and the electron 
EDM increase with $\sin(\phi_{\rm M})$, the electron electric dipole moment 
$d_e$ provides an important constraint on the realization of this 
electroweak baryogenesis scenario.  

For non-vanishing phases in the gaugino sector, the supersymmetric 
contribution to $d_e$ may become large and severe limits on the nMSSM 
parameter space can be obtained. Figure \ref{fg:edm} demonstrates that most 
of the LHC scan, for which $\phi_{\rm M}$ deviates substantially from zero 
or $\pi$, is excluded by the present 2 sigma upper limit $|d_e| < 1.9 \times 
10^{-27}$ e\,cm.  Since neither the LHC nor the ILC will detect the first 
generation sleptons if their masses are large, we allowed these masses to 
vary in the scans in a wide range: $1 < M_{e_R} < 10 {\rm ~TeV}, ~ 2.5 < 
M_{L_1} < 10 {\rm ~TeV}$.\footnote{Barring accidental cancellations between
different contributions, smaller values are only allowed by the present EDM
limits if the phase $\phi_{\rm M}$ is very small. In principle this case could
not be excluded by LHC data alone, since neither the slepton masses nor the
CP-violating phase can be well constrained. However, as can be seen in
figure \ref{fg:edm}, even after imposing these ad-hoc bounds on the slepton
masses, the LHC
data would allow nMSSM scenarios for which the value of the electron EDM
varies over several orders of magnitude. Therefore the parameter space scan
presented here should give a representative picture of the information that may
be obtained after running the LHC.}
For the LHC only those models survive the $|d_e|$ 
limit which either have small values of $\phi_{\rm M}$, very large values 
of the slepton masses, or where the one and two loop contributions to $d_e$ 
accidentally cancel.  Unfortunately, since this cancellation can happen at 
any value of $\phi_{\rm M}$, the EDM limit combined with the LHC data 
cannot shed light on the actual value of the phase $\phi_{\rm M}$.

New experiments have been proposed which are expected to improve the 
electron EDM limits by orders of magnitude in the next few years 
\cite{Lamoreaux:2001hb, Semertzidis:2004uu}.  If baryogenesis is driven by a 
single gaugino phase of the nMSSM such as studied in this work, then a 
non-vanishing value of $d_e$ will probably be measured by the time of the 
ILC operation as scenario A suggests.  This can even happen if the first 
generation sleptons are very heavy, as shown by the case of the input model 
A, where the first generation sleptons are fixed at $O(10 TeV)$.  If an 
electron EDM is measured, the indirect ILC determination of $d_e$ will also 
be an essential cross check confirming the nMSSM.  According to Figure 
\ref{fg:edm}, the ILC data alone will be able to constrain $|\phi_{\rm M}|$ 
below about 0.3 and $|d_e|$ below about $5 \times 10^{-28}$ e\,cm.  The 
correlation between $|d_e|$ and $|\phi_{\rm M}|$ is strong in the relevant 
$\phi_{\rm M} \sim$ 0.1 region. Utilizing this, even a 50\% measurement of 
$|d_e|$ combined with the ILC data will be able to constrain $\phi_{\rm M}$ 
in a $\pm 0.15$ window.

\subsubsection{Comment about parameter space for CP violation}
\label{sc:comment}

In the previous chapters we have made the simplifying assumption that the only
CP-violating parameter beyond the CKM matrix is a common phase of the gaugino
mass parameters
$\phi_{\rm M} \equiv \phi_{\rm M_1} = \phi_{\rm M_2} =\phi_{\rm M_3}$.
However, as pointed out in chapter \ref{sc:nmssm}, the general nMSSM with minimal flavor
violation can have
13 complex CP-violating parameters in total, four of which can be relevant for
electroweak baryogenesis. These four independent parameters can be chosen
to be $M_1$, $M_2$, $a_\lambda$ and $A_{\rm t}$.
For illustration, here we list the different scenarios that are possible if 
these phases are allowed to be non-zero. A more detailed investigation of these
scenarios would be beyond the scope of this paper and has to be performed
elsewhere.

\emph{Gaugino phases without gaugino unification:}
In this case, the baryogenesis could be driven by bino
currents, which are sensitive to the phase of $M_1$, while the phase of $M_2$
could be zero. While, depending on the mass pattern, this still could allow the
generation of a sufficiently large baryon asymmetry \cite{XXXX}, 
it would reduce the
contribution to the EDMs. As a result, lower values of the slepton masses of
the order of a few TeV would be allowed.
The collider sensitivity to $\phi_{\rm M_1}$ would be similarly weak as for the
common gaugino phase $\phi_{\rm M}$ discussed above, since one would rely on
the same observables in the neutralino sector.
Contrary to the MSSM, a phase in $M_1$ will have no sizeable effect on the dark
matter annihilation, since in the nMSSM the LSP is expected to be mainly
singlino.

\emph{Only CP-violating phase in $a_\lambda$:}
As shown in Ref.~\cite{huberco}, a sizeable phase in the Higgs sector could
succesfully explain electroweak baryogenesis. With the freedom of R- and
PQ-transformations, this phase could be attributed alternatively to $a_\lambda$,
$\tad$ or $m_{12}$, without changing the physics. In this situation, the
EDM constraints from one-loop diagrams will be much weaker than for a gaugino
phase, since one-loop diagrams with a Higgs boson are suppressed by an
additional small Yukawa coupling. As a result, in this scenario, light slepton
masses of a few hundred GeV would be allowed, which could be in reach of the
LHC or ILC. This feature could be used to experimentally discriminate this
scenario from a scenario with non-zero gaugino phases.
For instance, if the LHC or ILC would discover selectrons, the present electron
EDM limit would rule out phases in the gaugino sector that are large enough to
explain electroweak baryogenesis (except for the possiblity of large
cancellations between two contributions to the EDM).
A CP-violating phase in the Higgs sector could lead to observable effects in
Higgs production processes and decays. For example, mixing between CP-even and
CP-odd
Higgs states would enable the possibility to produce a third light Higgs boson
$\phi_i$ in the process $e^+ e^- \to \phi_i Z$. A more conclusive answer to
this question would require further study.
As pointed out in section~\ref{sc:direct}, a complex phase of $a_\lambda$ can
also strongly affect the dark matter scattering cross sections and thus could
relax the constraints coming from dark matter direct detection experiments.

\emph{Only CP-violating phase in $A_{\rm t}$:}
If the only non-zero phase (in addition to the CKM matrix) was in the parameter
$A_{\rm t}$, it is unlikely that a sufficiently large baryon asymmetry could be
generated, as experience from the MSSM shows \cite{X,XX,XXX,XXXX}. It is
however possible that a non-zero phase of $A_{\rm t}$ can exist in conjunction
with other complex nMSSM parameters.

\emph{CP-violating phases in several parameters:}
When considering to most general scenario with CP-violating phases in multiple
parameters, the interpretation of the EDM limits becomes more involved. From
our analysis, the prospects for disentangling the complex parameters through
collider data do not seem promising.

\section{Conclusions}
\label{sc:concl}

In this article, we have presented the phenomenological properties of the
nMSSM, assuming the parameters to be close to the ones that lead to a
solution of the dark matter and baryogenesis problems. A light neutralino
and chargino spectrum appears under these conditions, that can be probed
at the LHC and the ILC. 
At the LHC, however the detection of these weakly interacting
particles becomes difficult, unless they are produced from the cascade
decay of strongly interacting particles. In order to study the LHC
phenomenology we have therefore assumed that the gluino is within the reach of
the LHC, as suggested by gaugino mass unification, and that the
third generation squarks are light, which is helpful in avoiding the
suppression of the chiral charges
necessary for baryogenesis. We have shown that, under these conditions
a relatively good determination of the LSP and other neutralino 
masses may be obtained. However, the accuracy of these measurements is not
sufficient to allow the computation of the neutralino relic
density with any meaningful precision. A definitive
probe of this model can only be done at the ILC. At the ILC, for
a representative point, we have shown that both
chargino and four of the five neutralino masses may be determined.
The sparticle mass pattern leads to
a good discrimination of this model from the MSSM, and even a distinction
between the nMSSM and the NMSSM, which has a different singlet sector, is
possible. Moreover, the mass measurements, together with the production cross
section information, lead to
a good discrimination of this model from the MSSM, and allow to
compute reliably the annihilation cross section of the LSP,
and thus to check the agreement needed for a successful
explanation of the dark matter relic density.

The requirement of a sufficiently strong first order electroweak phase
transition also translates in this model into the presence of a light singlet
scalar, which mixes after electroweak symmetry breaking with the MSSM
Higgs doublets. Consequently, two light CP-even Higgs scalars appear
in the spectrum which couple with reduced couplings to the weak gauge
bosons, and which decay predominantly into the LSP. These CP-even Higgs
bosons may be searched for at the LHC and the ILC in the invisible decay
channel. At the LHC, however, it will be difficult to discern between one or
more invisibly decaying Higgs bosons. A definitive 
scrutiny of this question may only be performed at the ILC, at which
a good determination of the Higgs boson masses may be established.
A light CP-odd Higgs boson also appears in the spectrum, of difficult
detectability due to its strongly reduced couplings to the SM fermions.
Nevertheless, the pattern of the two CP-even Higgs boson masses allows to
distinguish the nMSSM Higgs sector from the MSSM. In addition, the mass
measurements provide information about the parameters of the model relevant for
a strong first order electroweak phase transition.
Note that for this kind of analysis, we have shown that the large
radiative corrections in the Higgs sector need to be under control, requiring
some information about  the superpartners of the top quark needs from
measurements at LHC or ILC.

Finally, we have also investigated the constraints coming from the  current
direct detection searches for dark matter, as well as the prospects of a
successful observation of the dark matter candidate of this model at future
experiments. We have shown that the predominantly singlet component of the LSP
makes its direct detection easier than in the MSSM  for the spin independent
channel, but more difficult for the spin depedent channel. For the spin
independent case, the current and next generation experiments should be able to
definitely probe this scenario. Similarly, assuming that the dominant phases
are in the chargino sector, we have also investigated the bounds coming from
the electron electric dipole moment. We have shown that, due to the small
values of $\tan\beta$ necessary to  realize this model, the EDM constraints
become weaker than in  the MSSM, and, again, this model may only be probed by
next generation experiments. If the dominant phases are in the Higgs sector
instead, the EDM bounds are even weaker. Such complex phases may also
affect the predicted dark matter direct detection
cross sections significantly. In this case
additional information about CP-violation might be obtainable from collider
data in the Higgs sector.

In summary, the nMSSM provides an exciting framework for addressing the
problems of baryogenesis and dark matter. The properties of this model and
role it plays in solving these problems could be probed with high precision at
the next generation of laboratory experiments, and would allow us to make
connections between laboratory and astrophysical observations at a new level of
insight.

\bigskip

\vspace{- .3 cm}

\section*{Acknowledgments}

We would like to thank T.~Barklow, J.~Hewett, A.~Menon, D.~Morrissey and
T.~Rizzo for useful communications and 
advice. We are grateful to V.~Barger and D.~Morrissey for discussions that 
led to an improvement of our direct detection calculations.
Work at ANL is supported in part by 
the US DOE, Division of HEP, Contract DE-AC-02-06CH11357, while work at
Universit\"at Z\"urich is partially supported by the Schweizer Nationalfonds.
Fermilab is 
operated by Universities Research Association Inc. under contract no. 
DE-AC-02-76CH02000 with the DOE. We gratefully acknowledge the use of {\it 
Jazz}, a 350-node computing cluster operated by the Mathematics and Computer 
Science Division at ANL as part of its Laboratory Computing Resource Center.
M.~C. and C.~W. are thankful to the Aspen Center for Physics, where part of this
work has been performed.

\end{document}